\documentclass[referee]{raa}            

\usepackage{graphicx,times}             
\usepackage{natbib}
\usepackage{amssymb,amsmath}
\usepackage{lscape}
\usepackage{pdflscape}
\usepackage{color}
\bibpunct{(}{)}{;}{a}{}{,}

\usepackage[pass,a4paper]{geometry}

\begin{document}

   \title{Stellar population analysis on the stacked spectra of double-peaked emission-line galaxies
}

   \volnopage{Vol.0 (20xx) No.0, 000--000}      
   \setcounter{page}{1}          

   \author{Meng-Xin Wang
      \inst{1,2}
   \and A-Li Luo
      \inst{1,2,3}
   }

   \institute{National Astronomical Observatories, Chinese Academy of Sciences,
             Beijing 100012, China; {\it lal@nao.cas.cn}\\
        \and
             University of Chinese Academy of Sciences, Beijing 100049, China\\
         \and
             Department of Physics and Astronomy, University of Delaware, Newark, DE 19716, USA\\
\vs\no
   {\small Received~~20xx month day; accepted~~20xx~~month day}}

\abstract{The double-peaked emission-line galaxies has long been perceived as objects related with merging galaxies or other phenomena of disturbed dynamical activities, such as outflows and disk rotation. In order to find the connection between the unique activities happening in these objects and their stellar population physics, we study the stellar populations of the stacked spectra drawn from double-peaked emission-line galaxies in the Large Sky Area Multi-object Fiber Spectroscopic Telescope (LAMOST) Data Release 4 (DR4) and Sloan Digital Sky Survey (SDSS) Data Release 7 (DR7) database. We group the selected double-peaked emission-line objects into 10 different types of pairs based on the Baldwin-Phillips-Terlevich (BPT) diagnosis for each pair of blue-shifted and red-shifted components, and  then stack the spectra of each group for analysis. The software STARLIGHT is employed to fit each stacked spectra, and the contributions of stars at different ages and metallicities are quantified for subsequent comparative study and analysis. To highlight the commonness and uniqueness in these double-peaked emitting objects, we compare the population synthesis results of the stacked spectra of double-peaked emission-line galaxies with that of their counterpart reference samples appearing single-peaked emission features. The reference sample are also selected from LAMOST DR4 and SDSS DR7 databases, respectively. From the comparison results, we confirm the strong correlations between the stellar populations and its spectral classes, and find that the double-peaked emitting phenomena is more likely to occur in the `older' stellar environment and the subgroups hosting different BPT components will show obvious heterogeneous star formation history.
}
   \authorrunning{M.-X. Wang, A.-L. Luo}            
   \titlerunning{Stellar population analysis on double-peaked emission line galaxies}  

   \maketitle

%

%
\section{Introduction}           
\label{sect:intro}

In recent years, with the advent of large-scale spectroscopic surveys of galaxies, more efforts have been invested into the search for double-peaked narrow emission lines galaxies, which were usually considered to be associated with merging galaxies or even dual active galactic nuclei (AGNs) \citep[e.g.][]{2009ApJ...705L..76W,2010ApJ...708..427L,2010ApJ...716..866S,2012ApJS..201...31G}. As it has been proved in previous works \citep[e.g.][]{2009ApJ...705L..76W,2013MNRAS.429.2594B}, if two galaxies merge to kiloparsec (kpc) scale, and the adjacent narrow emission lines emitting are sensed by a single spectrograph slit or fiber, double-peaked narrow emission line features would appear in the spectra. Follow-up observations of the reported double-peaked emission-line candidates have revealed that the confirmation rate is low \citep[e.g.][]{2011ApJ...735...48S,2012ApJ...753...42C}, which seems to conflict with expectations inferred from the galaxies' merging rate and the triggering of gas accretion onto black holes during merging stage \citep[e.g.][]{2005MNRAS.361..776S,2005ApJ...625L..71H,2012ApJ...748L...7V}. Based on the observationally estimated major merger rates of galaxies, and various scaling relations on the properties of galaxies and their central massive black holes, \citet{2011ApJ...738...92Y} suggested a phenomenological model to estimate the number density of dual AGNs and its evolution. The confirmation of kpc-scale binary or dual AGN systems from the selected double-peaked candidates still verify the reliability of systematically searches based on double-peaked criteria \citep[e.g.][]{2013ApJ...762..110L,2015ApJ...806..219C}. For instance, \citet{2013ApJ...762..110L} confirmed the binary-AGN scenario for two targets with high-resolution optical and X-Ray imaging from a parent sample of 167 Type II AGNs with double-peaked narrow emission lines selected from the Sloan Digital Sky Survey (SDSS; \citet{2000AJ....120.1579Y}), and \citet{2015ApJ...806..219C} identified six dual AGNs and dual/offset AGNs using X-rays based on a parent sample of 340 double-peaked emission-line AGNs identified in SDSS. Past researches have also revealed that other processes not associated with merging galaxies or dual AGNs can also produce the phenomenon of narrow double-peaked emitting, such as the disturbed narrow emission line regions (NLRs) involving biconical outflows \citep[e.g.][]{1994AAS...185.2705G,2011ApJ...727...71F,2011ApJ...735...48S,2012ApJ...745...67F,2018MNRAS.473.2160N}, the rotation-dominated regions with disturbance or obscuration \citep[e.g.][]{2016ApJ...832...67N,2015ApJ...813..103M}, and the local interaction of radio jets with NLR clouds \citep[e.g.][]{2010ApJ...716..131R}. Using the spatially resolved information from long-slit data, \citet{2016ApJ...832...67N} pinpointed the the origin of NLRs for a complete sample of 71 Type II AGNs at {\it z} $<$ 0.1 with double-peaked features from SDSS, and proposed a kinematic classification technique to discuss the scenarios that cause double-peaked emission lines in nearby galaxies. Based on the kinematic classification scheme and the follow-up optical long-slit observations of a sample of 95 SDSS galaxies that have double-peaked narrow AGN emission lines, \citet{2018ApJ...867...66C} confirmed that the majority of double-peaked narrow AGN emission lines are associated with outflows, and eight of these targets are compelling dual AGN candidates, and galaxies with double-peaked narrow AGN emission lines occur in such galaxy mergers at least twice as often as typical active galaxies.

The spectra of galaxies hold useful information on the stellar ages, and more valuable, stellar metallicitiy distributions of their composed stellar populations. To comprehend the stellar populations of diverse galaxies, we need to retrieve the stellar compositions of a galaxy from its integrated spectrum, which it never be an easy task. Stellar population synthesis on galaxies is developed as a means to estimate the physical properties of galaxies. The first group pioneered the empirical population synthesis method \cite{faber}, in which the integrated light of a galaxy can be reproduced by a linear combination of elements with known characteristics, such as the spectra of individual stars or star clusters holding diverse ages and metellicities from the spectra library. The further group introduced the so-called evolutionary population synthesis methods relying mainly on models \citep[e.g.][]{1978ApJ...222...14T,1983ApJ...273..105B,1996ApJ...457..625C,2003MNRAS.344.1000B}, which should follow the time evolution of the stellar system, and the parameters considered include the stellar initial mass function (IMF), star formation rate and the chemical histories. As we have seen, the latter technique received more widespread adoption \citep[e.g.][]{1996A&A...311..425B,1999ApJ...525..144V,2004A&A...425..881L}. Some works have been emphasized on the investigation of stellar populations on star-forming (SF) and starburst galaxies using the optical data \citep[e.g.][]{1996MNRAS.278..965S,2003MNRAS.340...29C}, and some works have been dedicated to AGNs associated research, such as a series of stellar population analysis conducted on low-luminosity AGNs by Cid Fernandes et al.(2004, 2005). In recent years, thanks to the explosion of integral-field spectroscopic (IFS) surveys, the astronomers have the opportunity to fully characterise the properties of galaxies at different redshifts with better spatial resolution. \citet{2010MNRAS.408...97K} has presented a stellar population analysis of the absorption line strength maps for 48 early-type galaxies from the Spectrographic Areal Unit for Research on Optical Nebulae (SAURON, \citet{2007MNRAS.379..401E}) sample. They have estimated the simple stellar population-equivalent age, metallicity and abundance ratio [$\alpha$/Fe] over a two-dimensional field extending up to approximately one effective radius, with the help of properties of H$\beta$, Fe5015 and $Mg\,b\,$ measured in the Lick/IDS system. \citet{2018MNRAS.478.5491M} have studied the effects of the active nuclei on the star formation history of the host galaxies, by presenting spatially resolved stellar population age maps, average radial profiles and gradients for the first 62 AGNs observed with SDSS-IV Mapping Nearby Galaxies at Apache Point Observatory (MaNGA, \citet{2015ApJ...798....7B}). Based on the data from the Multi Unit Spectroscopic Explorer (MUSE, \citet{2015A&A...575A..75B}), \citet{2018MNRAS.479.2443V} have measured the low-mass stellar IMF, and a number of individual elemental abundances, as a function of radius in NGC 1399, which is the largest elliptical galaxy in the Fornax Cluster. As for the merging galaxies, with which the double-peaked emission-line galaxies probably relate, there already have existed some research on their stellar populations. \citet{2015A&A...579A..45B} have studied the impact of the interaction in the specific star formation and oxygen abundance on different galactic scales, using the optical IFS data from 103 nearby galaxies at different stages of the merging event, from close pairs to merger remnants provided by the Calar Alto Legacy Integral Field Area (CALIFA, \citet{2012A&A...538A...8S}) survey.

In the last decade, the flourishing of multi-fiber spectrographs makes it feasible to obtain very large amount of spectra, and a large sample of double-peaked emission-line galaxies has been established. \cite{2012ApJS..201...31G} has made a systematic search to build the largest sample of 3030 double-peaked emission line galaxies from SDSS Data
Release 7 (DR7; \citet{2009ApJS..182..543A}) database, and \cite{2019MNRAS.482.1889W} has also set up a sample of 325 objects displaying double-peaked narrow line features based on the Large Sky Area Multi-object Fiber Spectroscopic Telescope (LAMOST) DR4 data set. The present sample from large-scale spectroscopic surveys enable us to probe into the influence of these unique dynamical activities on the stellar populations of these objects, and learn more about the relationship between nuclear activity and the star formation history in the inner regions of AGNs, which is one crucial question in the research of galaxy formation and evolution. In this work, we will investigate the stellar population physics of the double-peaked emission-line galaxies, with the prospect of finding their commonality and individuality compared with counterpart sample that comprises single-peaked emission-line galaxies.
 
The outline of this article is as follows. Section 2 displays the selections and classifications of our sample, Section 3 describes the STARLIGHT software, Section 4 presents spectral synthesis results and analysis based on the sample, and a summary is given in section 5. 


\section{Sample selection and classification}
\label{sect:Sample}

Our aim is to study the stellar population physics of the double-peaked emission-line galaxies, which harbor unique nuclear activity, (i.e., dual AGNs) or disturbed NLRs related with outflows, inflows or rotating disk. In this work, the samples are selected from galaxies belonging to LAMOST DR4 \citep{2015RAA....15.1095L} and SDSS DR7 \citep{2009ApJS..182..543A} datasets. 
\cite{2012ApJS..201...31G} has selected 3030 galaxies with double-peaked line profiles in prominent narrow emission lines, such as H$\beta$, [O~{\sc iii}$\lambda\lambda$4959,5007, H$\alpha$ and [N~{\sc ii}]$\lambda\lambda$6548,6584 from SDSS DR7, while \cite{2012MNRAS.420.1217D} has presented a comprehensive atlas of stacked spectra of galaxies with high high signal-to-
noise ratio (S/N) and resolution (S/N $\simeq$ 132$-$4760 at $\triangle\lambda$ = $1\mathrm{\AA}$), which are classified by color, nuclear activity and star formation activity, from the same dataset SDSS DR7. The stacked spectrum atlas is available online \footnote{http://www.vo.elte.hu/compositeatlas}.\cite{2019MNRAS.482.1889W} has built a sample of 325 objects displaying double-peaked narrow emission line features based on LAMOST DR4 data set, while \cite{2018MNRAS.474.1873W} has also provided a classification of typical emission-line galaxies based spectral line features of 40182 galaxies from LAMOST data product and created a set of stacked spectra \footnote{http://sciwiki.lamost.org/downloads/wll} for diverse galaxy classes within LAMOST DR4. 

To learn more about the features of double-peaked emission-line galaxies, we need a set of reference sample of single-peaked emission-line galaxies. Considering that the stacked spectra published in \cite{2012MNRAS.420.1217D} and \cite{2018MNRAS.474.1873W} can represent the characteristics of typical emission-line galaxies and they are also drawn from the same sources (i.e. SDSS DR7 and LAMOST DR4 database), we consider them as the reference sample. Limited by current data samples, we are unable to construct a particularly detailed reference sample, such as the sample with same stellar mass or similar redshift intervals, the ones defined in this work focus on Baldwin-Phillips-Terlevich (BPT) diagnosis (i.e. the emission-line flux ratio diagnostic diagram). As described in their works, according to the BPT diagnosis \citep{1981PASP...93....5B}, the stacked spectra of galaxies published by \cite{2012MNRAS.420.1217D} and \cite{2018MNRAS.474.1873W} are classified into SF galaxies, composite galaxies (the one falls into the transition region between SF and AGN, hereafter denoted as AGN + [H~{\sc ii}] for being able to distinguish from composite spectra), Seyfert galaxies and LINERs. Likewise, to get a general view of the stellar populations of double-peaked emission-line galaxies in different spectral class and to improve the S/N of the spectra, we perform the stellar population synthesis fittings on the called stacked spectra, which are combined from double-peaked emission-line galaxies from \cite{2012ApJS..201...31G} and \cite{2019MNRAS.482.1889W} and can be seen as the representative spectra for different spectral classes. The spectral classes are divided based on locations of each galaxy spectrum's blue-shifted and red-shifted components in BPT diagram, and this will be described in more detail later in this section.
 We compare the stellar population synthesis results between the representative stacked spectra from double-peaked objects with their counterpart stacked typical emission-line spectra, both of which are classified based on different levels of activity, to give insight to the similarities and dissimilarities between the two ones.

\begin{figure*}
\centering
\includegraphics[width=0.9\textwidth]{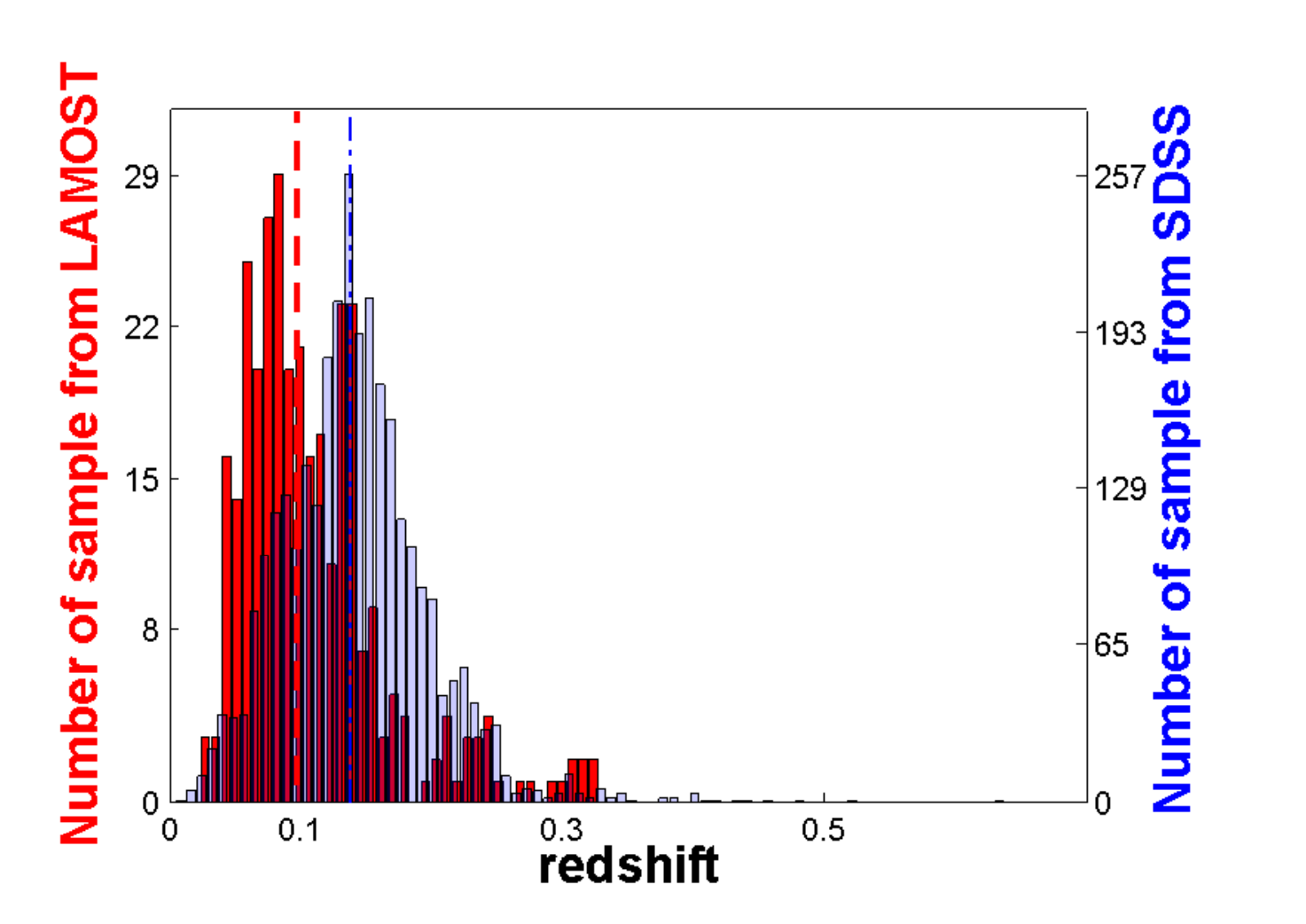}
\caption{The blue bars denote the distributions of redshifts for 3030 samples of \cite{2012ApJS..201...31G} selected from SDSS DR7 database, and the red bars represent the distribution of redshifts for 325 samples of \cite{2019MNRAS.482.1889W} selected from LAMOST DR4 database. The blue dash-dot line and red dashed line stand for the median redshifts of samples from SDSS and LAMOST, respectively.}
\label{fig1}
\end{figure*}

As we know, stellar populations synthesis calls for high accurate continua of spectra. However, some uncertainties ($<$10\%) are embedded in the shape of continuum of LAMOST spectra \citep[see][]{2016ApJS..227...27D}, mainly due to its current adopted method of relative flux calibration, which may bring about an imprecise outputs of spectral fitting \citep{2015RAA....15.1095L}. The extinction uncertainties existed in standard stars (i.e., high-quality $F$ dawrfs) may lead to errors in the deduced response curve and thus result in some uncertainty in flux calibration, and past analysis has also revealed that the lower the latitude in which the object lies, the greater the uncertainty would exists in its response curve \citep{2018MNRAS.474.1873W}. To obtain more accurate stellar populations, we adopt the method proposed in \cite{2018MNRAS.474.1873W} to eliminate the color inaccuracy in LAMOST spectra introduced by relative flux calibration. The method calibrates the original flux of each LAMOST spectrum using a low-order polynomial, which is derived from the fitting between LAMOST spectral flux and the converted flux constructed from $g$, $r$ and $i$ fibre magnitudes of its cross-referenced SDSS counterpart's photometric catalogue. As for the host redshift, which is of vital importance for line flux determination and stellar population analysis, we take it directly from catalogues of \cite{2012ApJS..201...31G} and \cite{2019MNRAS.482.1889W}. The redshifts are derived from direct pixel-fitting of the stellar absorption lines with templates (i.e., the Indo-U.S. Library of Coud{\'e} Feed Stellar Spectra from \cite{2004ApJS..152..251V}), and show a reasonable uncertainty of $\sim$20 km s$^{-1}$. FIGURE \ref{fig1} presents the distribution of redshifts of the two samples. The redshifts of the Ge's samples locate between 0.0082 and 0.6352, with a median of 0.1379, while for samples from \cite{2019MNRAS.482.1889W}, the redshifts rang from 0.0271 to 0.3246, with a median value 0.0968.
 
\begin{figure}
\centering
\includegraphics[width=0.9\textwidth]{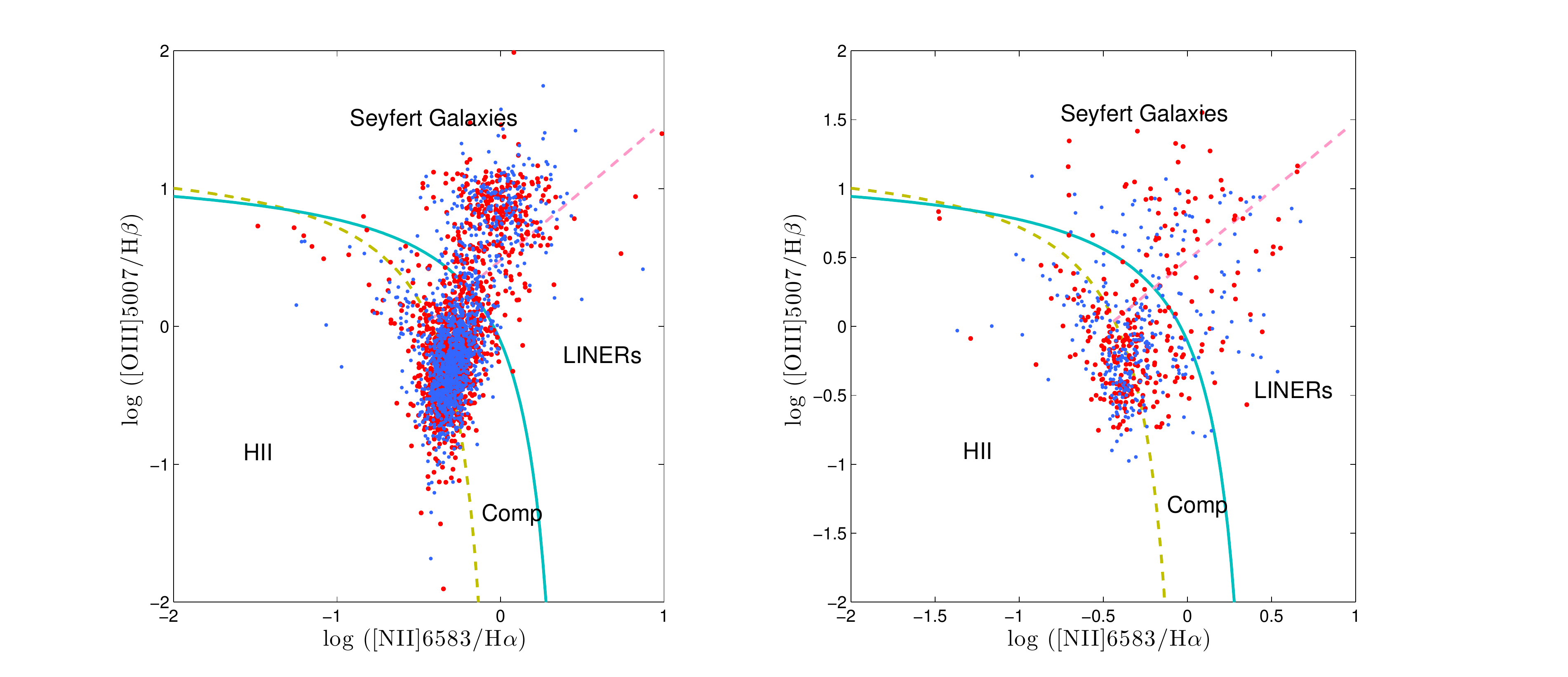}
\caption{BPT diagrams of blue and red components of double-peaked emission-line sample from SDSS DR7 database (left panel) and LAMOSR DR4 database (right panel). In both panels, the green solid and yellow dotted lines represent the classification lines suggested by \citet{2001ApJ...556..121K}, and \citet{2003MNRAS.346.1055K}, respectively, the pink dashed lines represent the locus defined by \citet{2010MNRAS.403.1036C} for Seyfert/LINER division. The blue symbol represents blue-shifted component, and the red symbol corresponds to red-shifted component, of each selected target.}
\label{fig2}
\end{figure}

The stacked spectra of typical emission-line galaxies provided by \cite{2012MNRAS.420.1217D} and \cite{2018MNRAS.474.1873W} are representatives of SF galaxies, AGN + [H~{\sc ii}], LINERs and Seyfert 2s based on BPT diagram with classification lines suggested by \citet{2001ApJ...556..121K} and \citet{2003MNRAS.346.1055K}. In \cite{2012MNRAS.420.1217D}, the stacked spectra are constructed using the principal component analysis (PCA) based method, while in \cite{2018MNRAS.474.1873W}, the stacked spectra are developed by the median calculation.
To make a one-to-one comparison, we also classify the double-peaked emission-line samples selected from SDSS and LAMOST in a similar way. Here we refer to the double-peaked emission-line profiles as multi-components composed of three kinematic groups, including the red-shifted and blue-shifted narrow emission lines (H$\beta$, [O~{\sc iii}]$\lambda\lambda$4959,5007, H$\alpha$ and [N~{\sc ii}]$\lambda\lambda$6548,6584), the [O~{\sc iii}] wings and broad Balmer emission lines. The multi-gaussians fitting is conducted to remodel each spectrum and the fluxes of the blue-shifted and red-shifted narrow components are derived from the fitting parameters. As the blue-shifted and red-shifted narrow components can be created by different ionization mechanisms, and the nature of each component can be diagnosed by examining its location in BPT diagram. FIGURE \ref{fig2} shows the final samples in BPT diagram, with the red-shifted and blue-shifted components being illustrated by different colors and symbols respectively, the AGNs and SF galaxies dividing lines developed by \citet{2001ApJ...556..121K} and \citet{2003MNRAS.346.1055K} are plotted as well. To distinguish Seyfert 2s from LINERs, an alternative dividing line from \citet{2010MNRAS.403.1036C} is adopted, which proposes a more economical and simpler Seyfert/LINER division diagnosis than \citet{2006MNRAS.372..961K}. As done in Ge's catalogue, the classification and statistic analysis are based on the looser Kauffmann's criteria, which takes the galaxies fall into the transition region as type II AGNs. Here we reprocess and reclassify the spectra of 1945 narrow emission-line objects which are marked as `2 type II', `type II+SF' and `2 SF' in his catalogue. Finally we make a new reclassification for 1473 objects from Ge's catalogue and 472 spectra are discarded owning to their high noise and non-ideal spectral multi-guassians fittings. For catalogue from \cite{2019MNRAS.482.1889W}, since there exist 18 type I AGN and 11 ``unknown" type objects as it has been stated in that work, we use the rest 296 objects for analysis. As the blue-shifted and red-shifted components for each double peaks could be driven by different mechanisms, there exist several grouping combinations for the sample. In this work, the BPT types considered are SF galaxies, composite galaxies, Seyfert 2s and LINERs, thus we refer to the 10 combined subgroups as types `2-SF', `2-COM', `2-Seyfert 2s', `2-LINERs', `SF+COM', `SF+Seyfert 2s', `SF+LINERs', `Seyfert 2s+COM', `COM+LINERs', and `Seyfert 2s+LINERs'. Here, as an instance, and so forth, the type `2-SF' stands for the combined subgroup, in which each galaxy consists of two shifted SF components, while the type `SF+Seyfert 2s' stands for the subgroup, in which each one consists of two shifted  components falling into the SF and Seyfert 2s regions on BPT diagram, respectively. TABLE \ref{number} shows the number of spectra within each combined subgroup for 1473 objects from SDSS and 296 objects from LAMOST, respectively. In the end, we combine all spectra within each subgroup into one stacked spectrum by interpolating them uniformly and calculating the median intensity at each pixel as it has been done in \cite{2018MNRAS.474.1873W}, and get the 10 representative spectra for double-peaked emission-line galaxies from LAMOST and SDSS, respectively. Here considering that the mean intensity calculated from each subgroup would be more vulnerable to poor spectra with high noise therein, we use the median method instead of a mean combination. As we see, the combination of two shifted components could affect the continua, and thus the spectral distribution of energy (SED) of double-peaked emission-line spectra will be different from their counterparts. Here we employ the software, called STARLIGHT \citep{2005MNRAS.358..363C}\footnote{http://www.starlight.ufsc.br}, to fit the spectral absorption lines and continua of the stacked spectra and to obtain their stellar populations. In this way, we can quantify the uniqueness of these double-peaked emitting objects.

\begin{landscape}
\begin{table*}
\caption{Number of Each Subsample.} 
\label{number} 
\centering
\begin{tabular}{r|c|c|c|c|c|c|c|c|c|c}
 \hline\hline   {Combined BPT Subgroup} &2-SF & 2-COM & 2-Seyfert 2s & 2-LINERs & SF+COM
&SF+Seyfert 2s&SF+LINERs&Seyfert 2s+COM&COM+LINERs&Seyfert 2s+LINERs\\
\hline
Sample from SDSS &305   & 478   & 208  & 12   & 302&7&3&59&43&56\\
\hline
Sample from LAMOST &96&48&28&10&60&2&8&11&15&18\\
\hline
\end{tabular}
\end{table*}
\end{landscape}

\section{Stellar Population Synthesis}
\label{sect:starlight}
STARLIGHT integrates the techniques initially developed for empirical population synthesis with the evolutionary synthesis models. Briefly speaking, an observed spectrum $O_{\lambda}$ is fitted with a model $M_{\lambda}$ that is a combination of $N_{\ast}$ simple stellar populations (SSPs) from the evolutionary synthesis models of BC03. The line-of-sight stellar motions are modeled by a Gaussian distribution $G$, with a central velocity $v_{\ast}$ and a dispersion $\sigma_{\ast}$. Extinction is represented as the parameter V-band extinction $A_{v}$, and the Galactic extinction law from Cardelli, Clayton, and Mathis (CCM, \citet{1988ApJ...329L..33C}) with $R_{v}=3.1$ is employed. In this work, we use a base of 45 SSPs, including 15 ages ranging between 1 $Myr$ and 13 $Gyr$ and three levels of metallicities 0.2 $Z_{\odot}$, $Z_{\odot}$ and 2.5 $Z_{\odot}$. These spectra are extracted from the Stellar
Library (STELIB, \cite{2003A&A...402..433L}), using the ``Padova-2004" models and an IMF from \cite{2003PASP..115..763C}. All these bases are normalized at wavelength $4020\mathrm{\AA}$. Prior to STARLIGHT, at the pre-processing stage, we complete the foreground extinction correction using the reddening maps from \cite{1998ApJ...500..525S}, bring all spectra to the rest frame based on the host redshifts, and sample them into a step of $1\mathrm{\AA}$ from wavelength $3701\mathrm{\AA}$ to $8500\mathrm{\AA}$. The masked band regions for STARLIGHT input are also compiled \citep[see][]{2005MNRAS.358..363C}. The regions include the bands around bad pixels, emission lines, sky lines, and some wavelength windows: $5870-5905\mathrm{\AA}$, to skip the $Na\,D\,\lambda\lambda5890,5896$ doublet; $6845-6945\mathrm{\AA}$ and $7550-7725\mathrm{\AA}$, the ranges covering the bugs listed by BC03 resulting from problems in the STELIB library \citep{2003A&A...402..433L}); $7165-7120\mathrm{\AA}$, which displays a systematic broad residual in emission as has been mentioned in \citet{2006MNRAS.370..721M} and $5800-6100\mathrm{\AA}$, which is partly located in the 200\AA ~ overlapped range (5700-5900\AA) between the red and blue band cameras of LAMOST. STARLIGHT directly provides the parameters of spectral synthesis, one important parameter is the population vector $x_{j} (j=1,...,N_{\ast})$, which reveals the contribution of each SSP to the model flux at the normalized wavelength $4020\mathrm{\AA}$.

\section{Synthesis Results}
\label{sect:results}

\subsection{Detailed Description of the Synthesis Results}
\label{sect:Robust Description}
We take the combined subgroups `2-COM', `SF+COM' ,`SF+Seyfert 2s' and `SF+LINERs' from SDSS sample as instance, to assess the reliability of stacked spectra as the representative ones for stellar population  analysis in this work. The subgroup `2-COM' is selected since it harbours the largest sample within these combined subgroups, as we can see from Table \ref{number} and each galaxy within it has the same BPT diagnosis for its blue-shifted and red-shifted emission-line components. The subgroup `SF+COM' is chosen to stand for types, whose dual shifted components appearing different BPT classifications. The subgroups `SF+Seyfert 2s' and `SF+LINERs' are also selected, as these two ones include only 7 and 3 objects respectively, and with these we can assess the robustness of stacked spectra in sparse samples. We perform stellar population synthesis on all spectra from the 4 representative subgroups, and get the fractional contributions of populations, within three different bins of ages and three levels of stellar metallicities, to light profile of each galaxy spectrum (these 6 types of quantified stellar populations will be described in detail in SUBSECTION \ref{sect:SDSS results}), then we calculate the mid-values of the contributing percentages for each subgroup, and list the results in TABLE \ref{s.3}. We compare the results with the counterpart values in TABLE \ref{s.ge}, which presents the percentages of light contributions to 10 stacked spectra constructed from different combined subgroups. We can see that for subgroups `2-COM', `SF+COM' and `SF+Seyfert 2s', the synthesis results displayed in TABLES \ref{s.3} and \ref{s.ge} show a similar proportional distribution. As for subgroup `SF+LINERs', in which there are only 3 spectra, it still shows a similar proportional distribution between results deduced from the stacked spectrum and all involved spectra, when we focus on the contributions from populations of different ages. While this is not the case when we consider the contributions from populations of different stellar metallicities. As we see, it is never be an easy task to generate an ideal stacked spectrum from such a small sample size, since in this case, the representative spectrum would be more susceptible to some spectrum with poor spectral quality when constructing. As posed above, and considering that the subgroup `SF+Seyfert 2s', which includes only 7 samples and still produces a reliable stacked spectrum, we can conclude that stacked spectrum can reflect the commonalities of samples within its subgroup in most cases. In our following analysis, for some subgroups (i.e., SF+Seyfert 2s from LAMOST, SF+LINERs from SDSS, see TABLE \ref{number}), which contain objects fewer than 7, we take the result as a reference.

\begin{table*}
\caption{The Mid-values of fractional contributions in 3 representative subgroups from \citet{2012ApJS..201...31G} of SDSS.}
\label{s.3} \centering
\begin{tabular}{c|c|c|c|c|c}
\hline\hline \multicolumn{2}{c|}{SSP}
  & \multicolumn{4}{c}{Emission-line Diagram}\\
\hline
  &  & 2-COM& SF+COM&SF+Seyfert 2s &SF+LINERs\\
\hline
age& young &	25.3	&	30.9	&27.0 &	16.4	\\
&intermediate&55.9	&	57.9	&44.8&	69.0	\\
&old&	11.8	&	6.4	&5.4&	8.5	\\	
&power law &  & &11.2 & 5.3\\
\hline
$Z/Z_{\odot}$ &0.2 &51.4	&	55.5	&61.1&	14.9\\
&1.0 &22.4	&	16.2	&11.7	&42.1\\
& 2.5&18.7	&	18.5	&6.2&	27.1	\\
&power law &  & &11.2& 5.3\\
\hline
\end{tabular}
\end{table*}

\subsection{Results And Comparisons Based On SDSS Data Set}
\label{sect:SDSS results}
At first, we conduct STARLIGHT on the stacked spectra from SDSS sample published in \cite{2012MNRAS.420.1217D}. FIGURES \ref{fig7} to \ref{fig10} in APPENDIX \ref{sect:appendix} show the synthesis outputs (the layout of each figure and its supplementary information will be explained in APPENDIX \ref{sect:appendix}). In this analysis, we use a not robust but detailed prescription of the star formation history of a galaxy by dividing the ages of stellar populations into three bins: `young' ($\leq5\times10^{8}yr$), `the intermediate-age' ($6.4\times10^{8}yr \leq age \leq5\times10^{9}yr$) and `old' ($\geq1\times10^{10}yr$), with the same criteria adopted in \cite{2005MNRAS.358..363C}. TABLE \ref{s.Dobos} presents the fractional contributions of populations within three quantified bins of ages and three levels of stellar metallicities to the model flux, respectively. From TABLE \ref{s.Dobos}, we can see that SF galaxies, AGN + [H~{\sc ii}] and Seyfert 2s hold abundant young and intermediate-age populations (from 92.3\% to 74.3\%), while LINERs contains a great deal of intermediate-age and old populations ($\sim$86\%). There also exists an obvious decrease of young population contribution from SF galaxies, AGN + [H~{\sc ii}], Seyfert 2s to LINERs, which is consistent with the conclusion in previous articles \citep[e.g.][]{2004A&A...428..373B,2009A&A...495..457C,2008MNRAS.391L..29S}, since the activities in Seyfert 2s and LINERs are perceived to be linked to nuclear, or possibly shocks for LINERs, while those in SF galaxies are dominated by stellar activities of young stars. Another conclusion we can draw from FIGURES \ref{fig7} to \ref{fig10} is that the intermediate-age and old populations contribute the vast majority to mass in all these subsamples, which also meets with the phenomena observed by \cite{2009A&A...495..457C}. As for the analysis of metallicity contributions, since we are considering three levels of metallicities, 0.2 $Z_{\odot}$, $Z_{\odot}$ and 2.5 $Z_{\odot}$, we calculate the percentages by adding up the contributions of relevant SSPs within each metallicity grid. It can be seen from TABLE \ref{s.Dobos} that SF galaxies and AGN + [H~{\sc ii}] are dominated by metal-poor population with metallicity 0.2 $Z_{\odot}$, while Seyfert 2s and LINERs posses a large amount of contribution from population with metallicity $Z_{\odot}$. We can see from TABLE \ref{s.Dobos} that LINERs show the most $Z_{\odot}$ contribution, the least 0.2 $Z_{\odot}$ contribution, the most `old' population and least `young' population contributions, which indicates its features of oldest, evolved metal-rich and little star formation activity, as has been summarized in \cite{2004A&A...428..373B}. A power law, which represents the non-stellar component \citep{1978ApJ...223...56K}, is also added for spectral fitting of LINERs and Seyfert 2s. The flux contributions of power-law spectrum in Seyfert 2s and LINERs are both little, less than $\sim$1\%, which is different from what has been concluded in past works and synthesis outputs on other stacked spectra in this work. This irrational output may attribute to PCA-based method used by \cite{2012MNRAS.420.1217D} when computing the average spectra, which is different from the traditional averaging and median calculation applied by us. The PCA-based method may ignore this featureless continuum component, and this needs to be further verified.

\begin{figure*}
\centering
\includegraphics[width=0.48\textwidth]{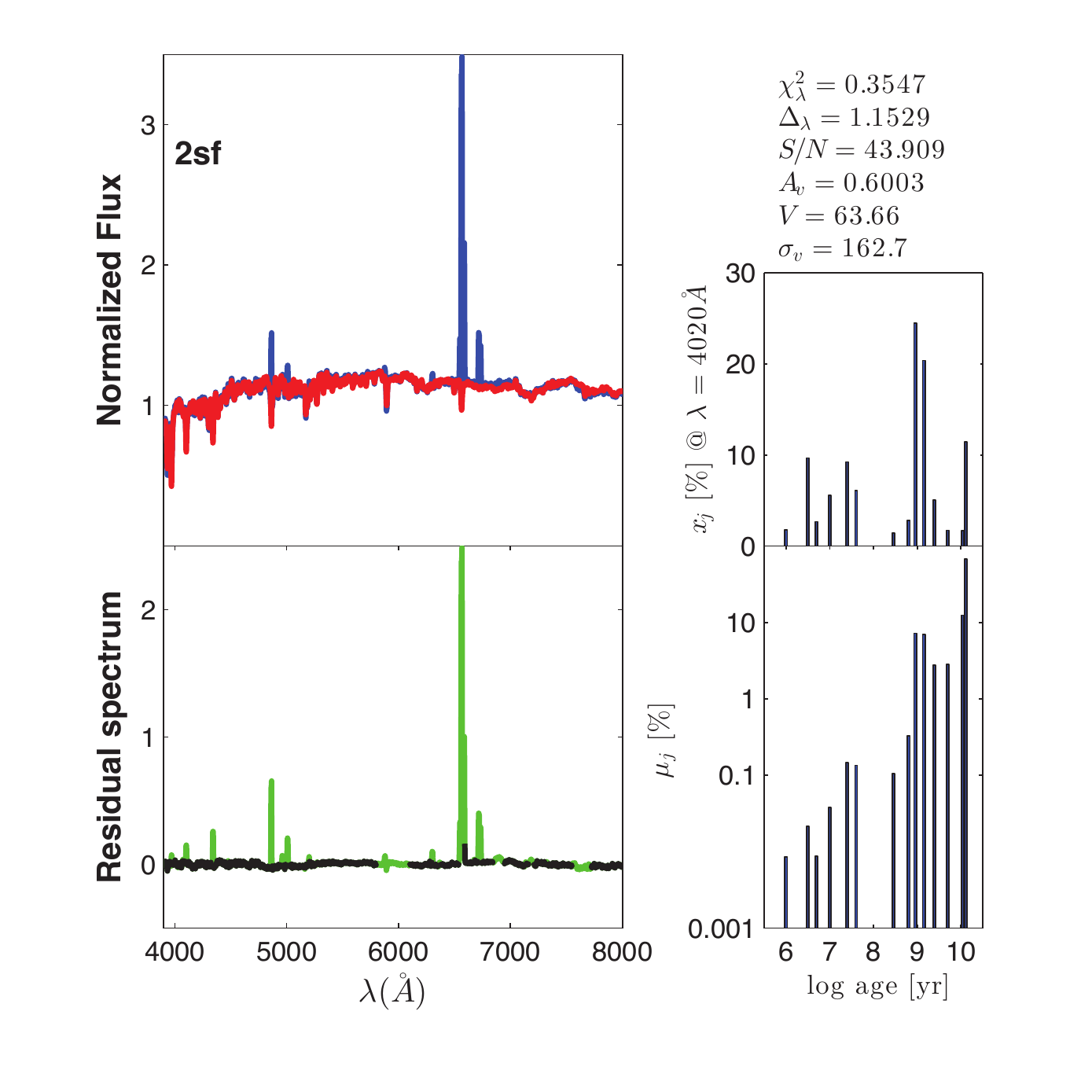}
\includegraphics[width=0.48\textwidth]{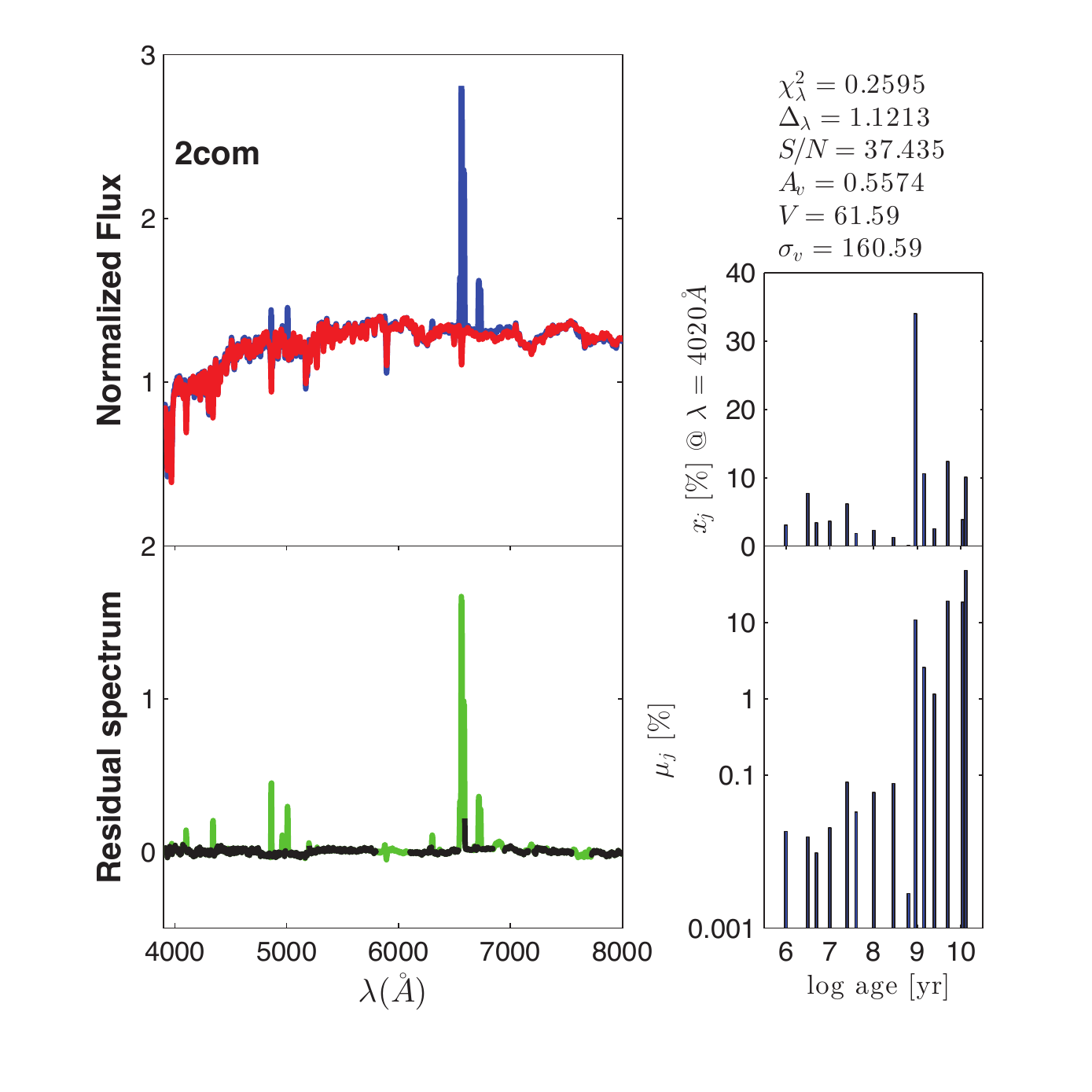}
\includegraphics[width=0.48\textwidth]{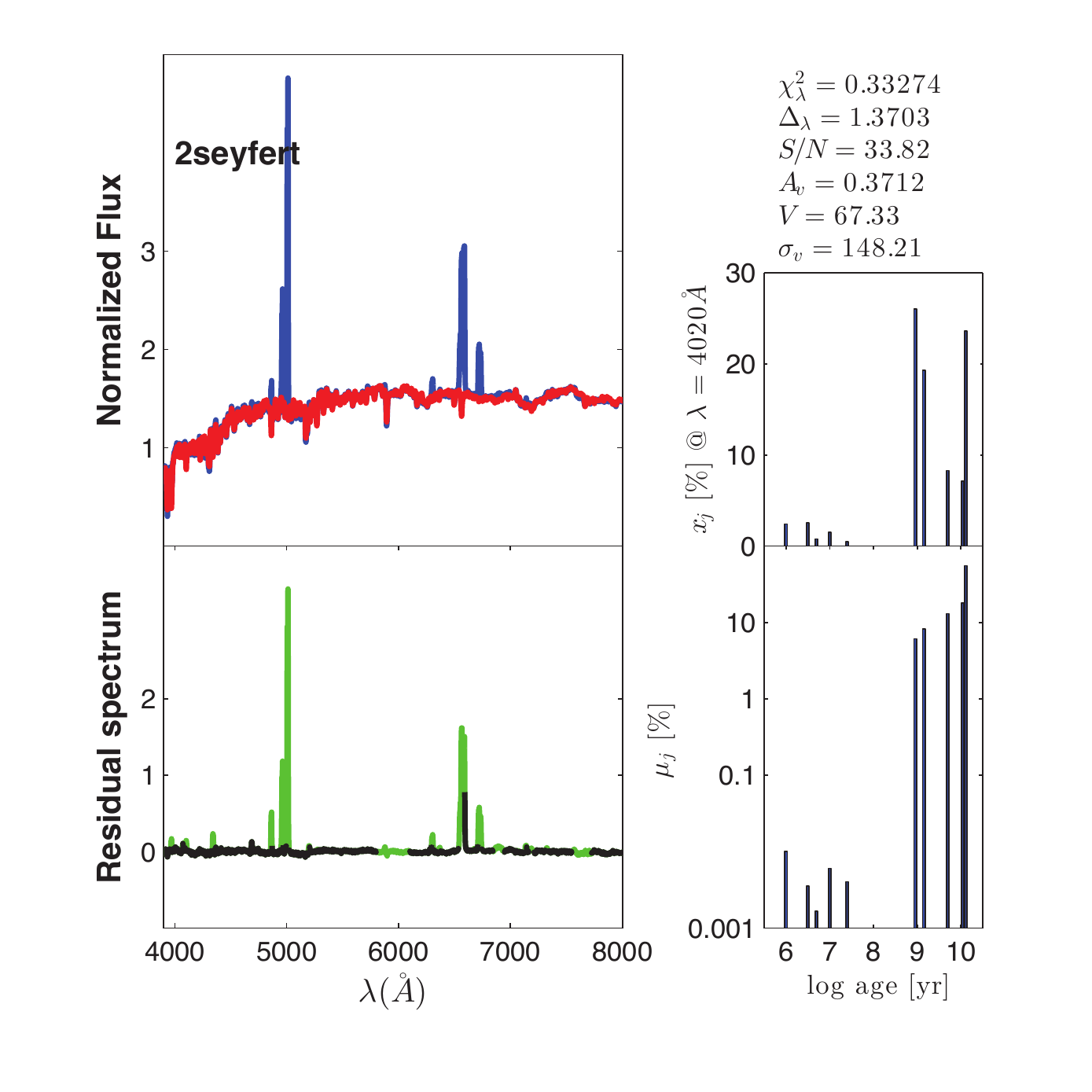}
\includegraphics[width=0.48\textwidth]{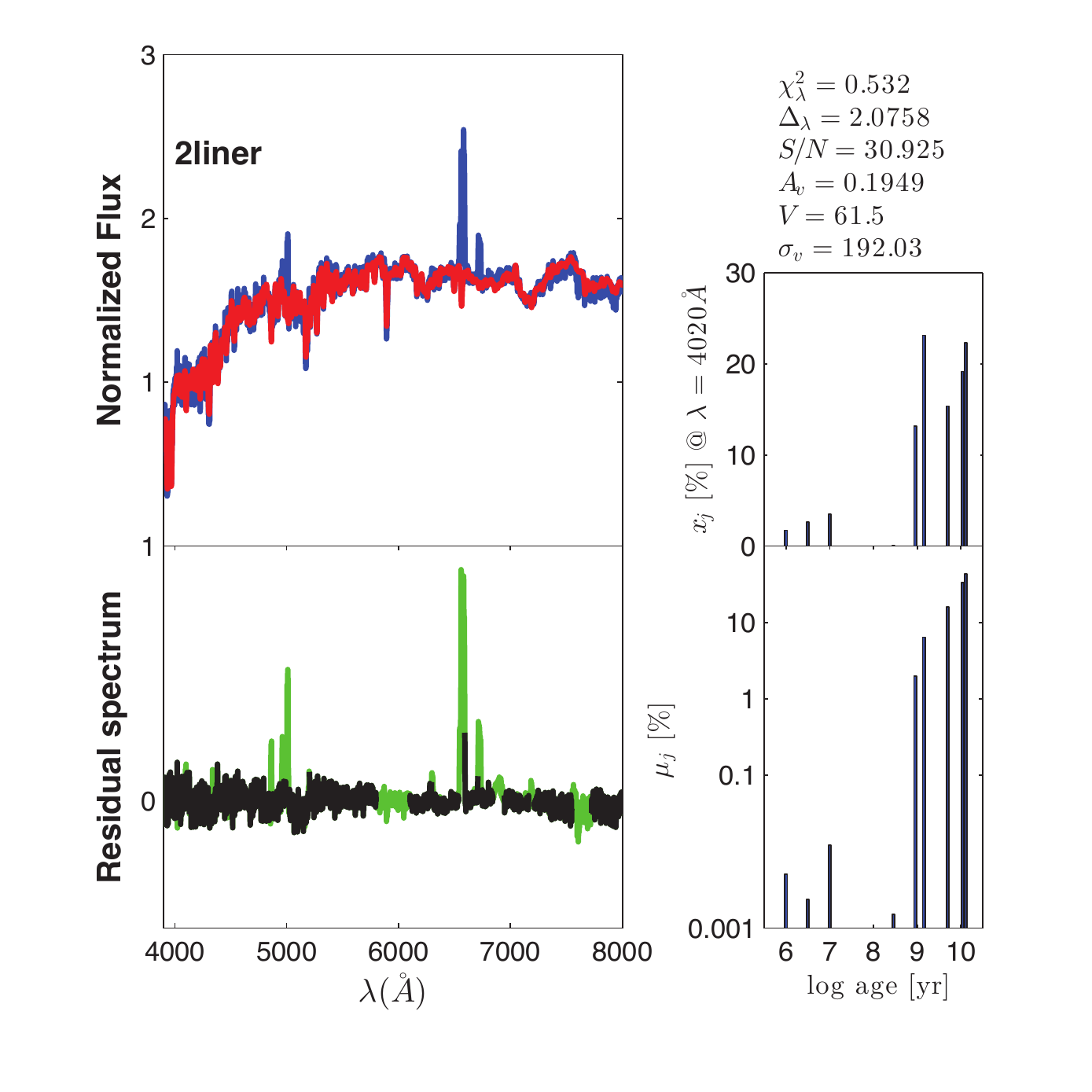}
\includegraphics[width=0.48\textwidth]{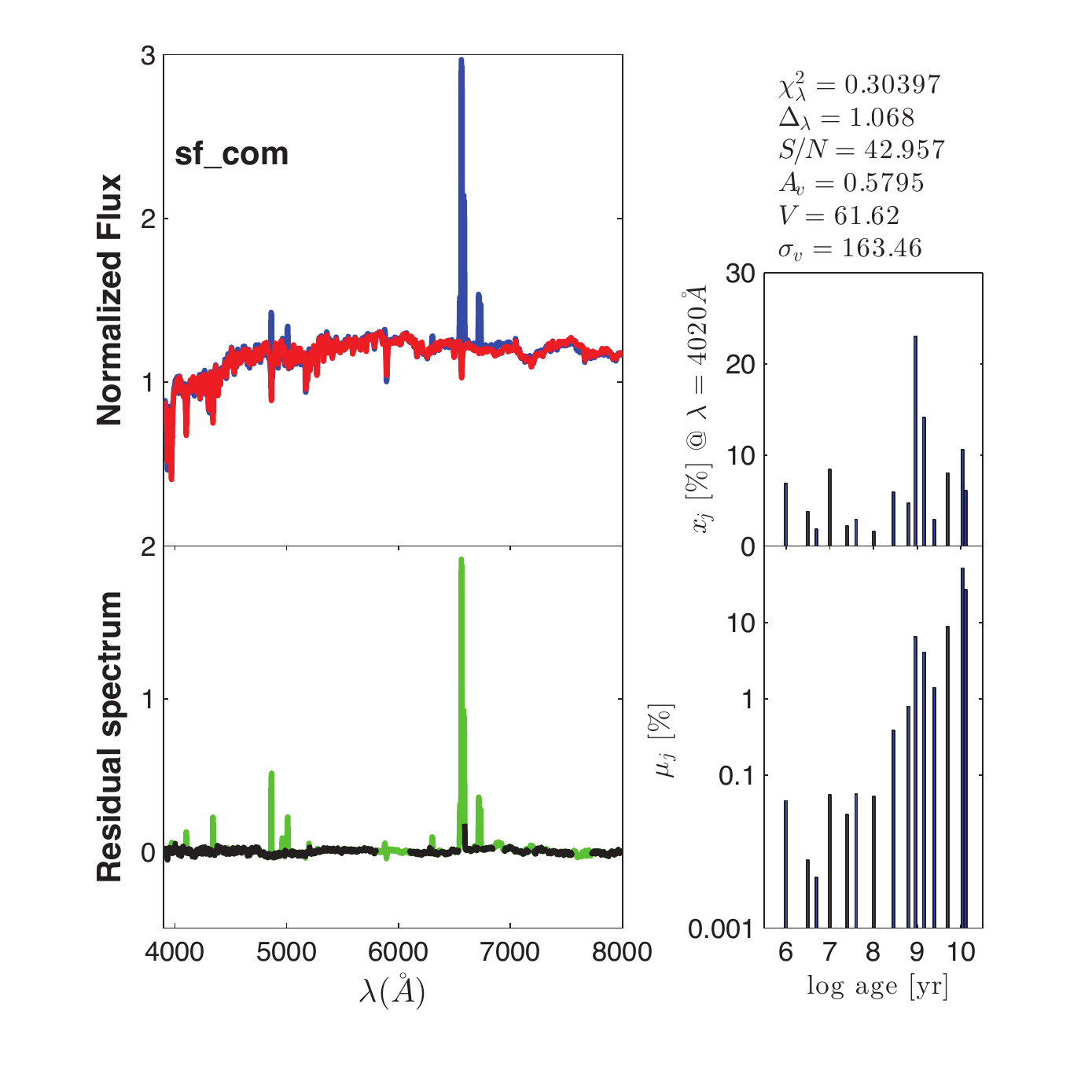}
\includegraphics[width=0.48\textwidth]{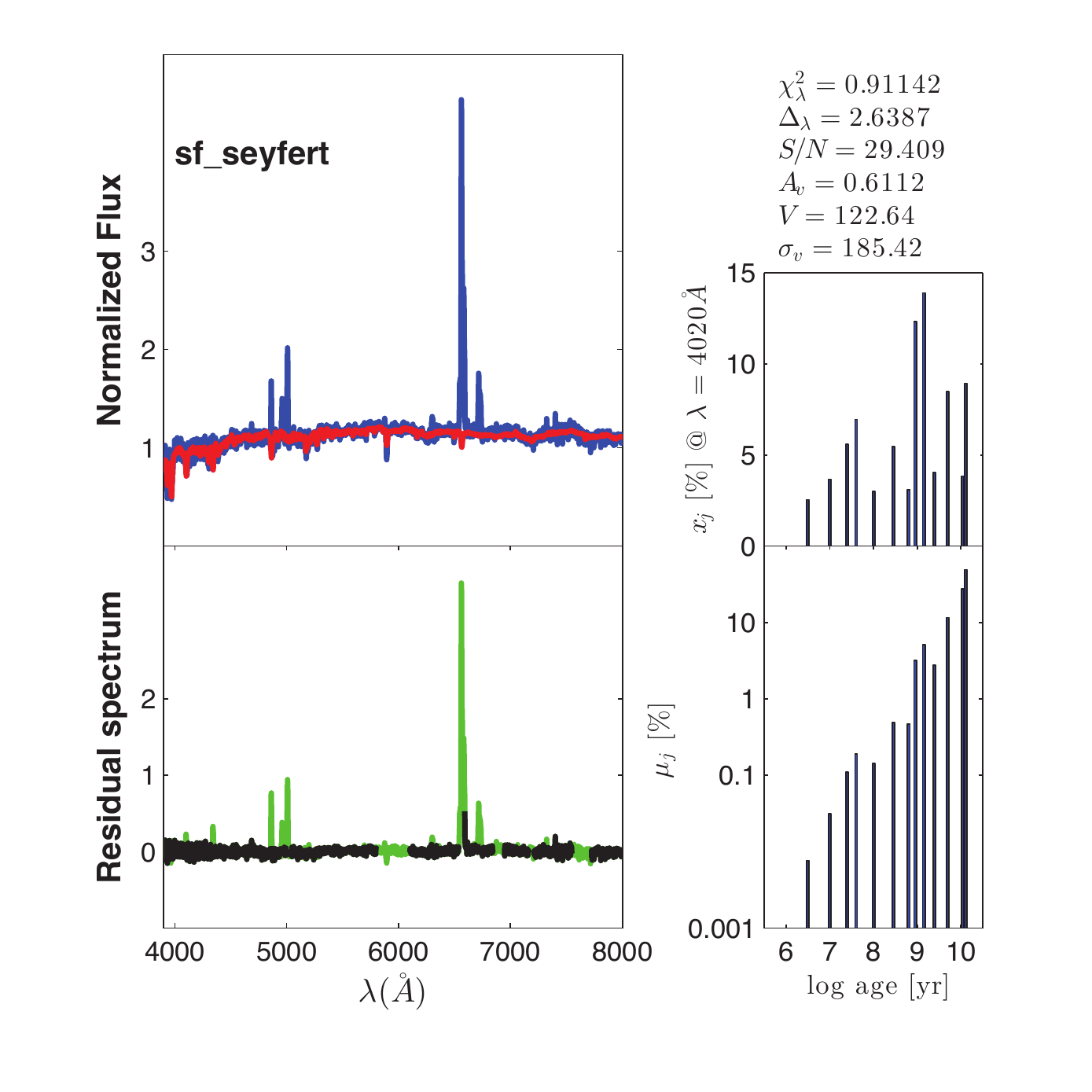}
\caption{Spectral synthesis outputs of 6 stacked spectra for double-peaked emission-line galaxies from SDSS sample published in \cite{2012ApJS..201...31G}. Each sub-figure shares the same layout as FIGURE \ref{fig7}, and types of the 6 stacked spectra are also tagged on top left of each sub-figure. All symbols are the same as in FIGURE \ref{fig7}.}
\label{fig3}
\end{figure*}

\begin{figure*}
\centering
\includegraphics[width=0.48\textwidth]{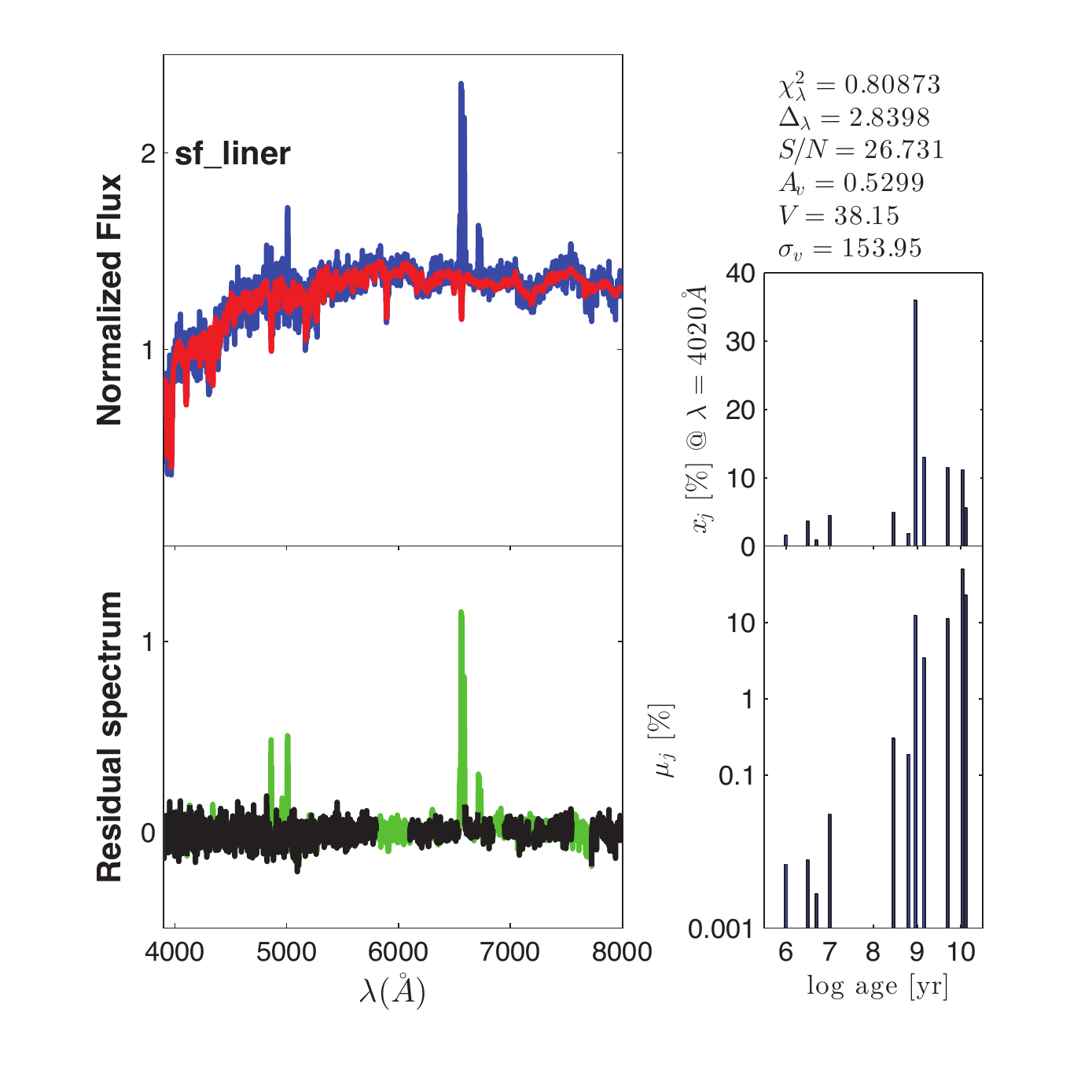}
\includegraphics[width=0.48\textwidth]{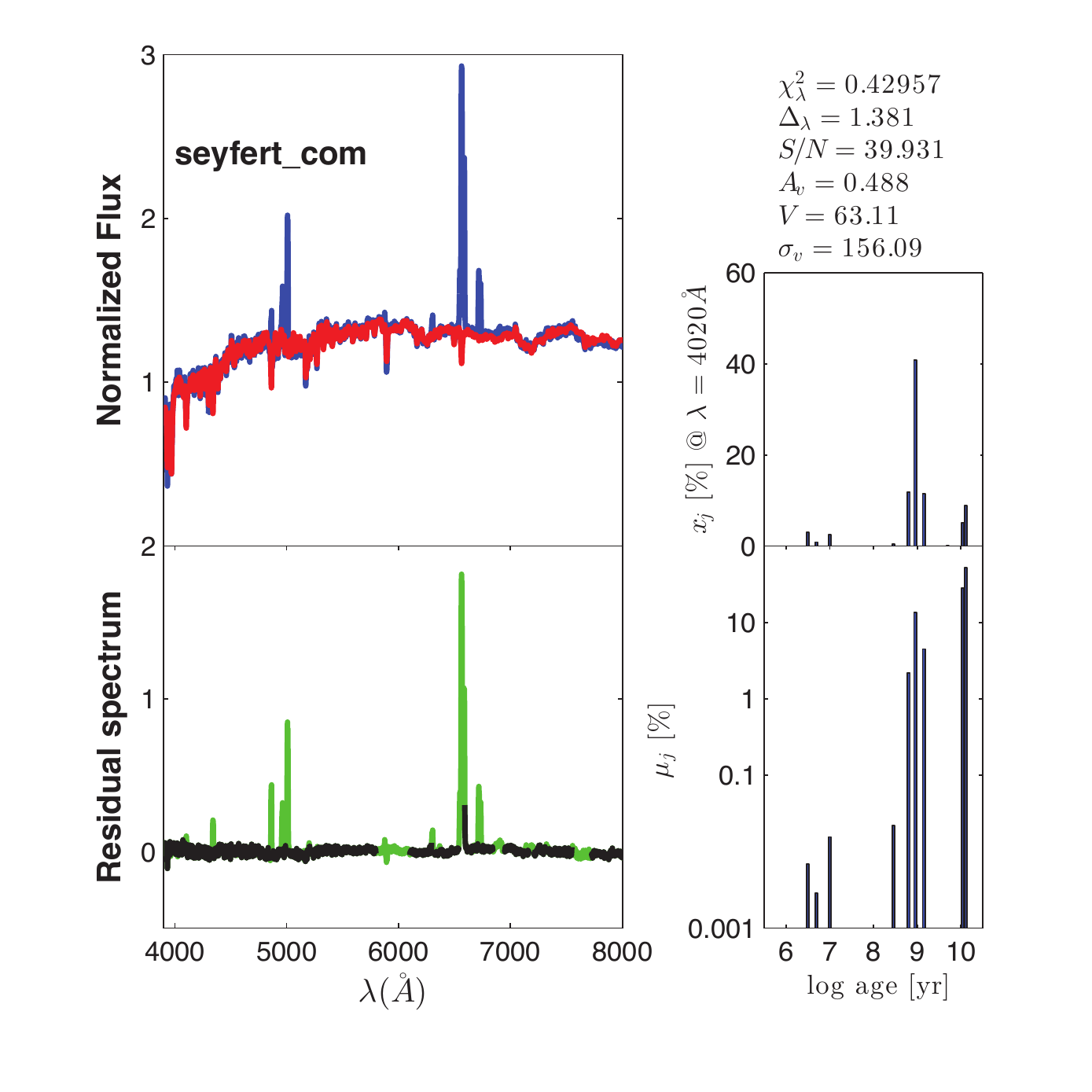}
\includegraphics[width=0.48\textwidth]{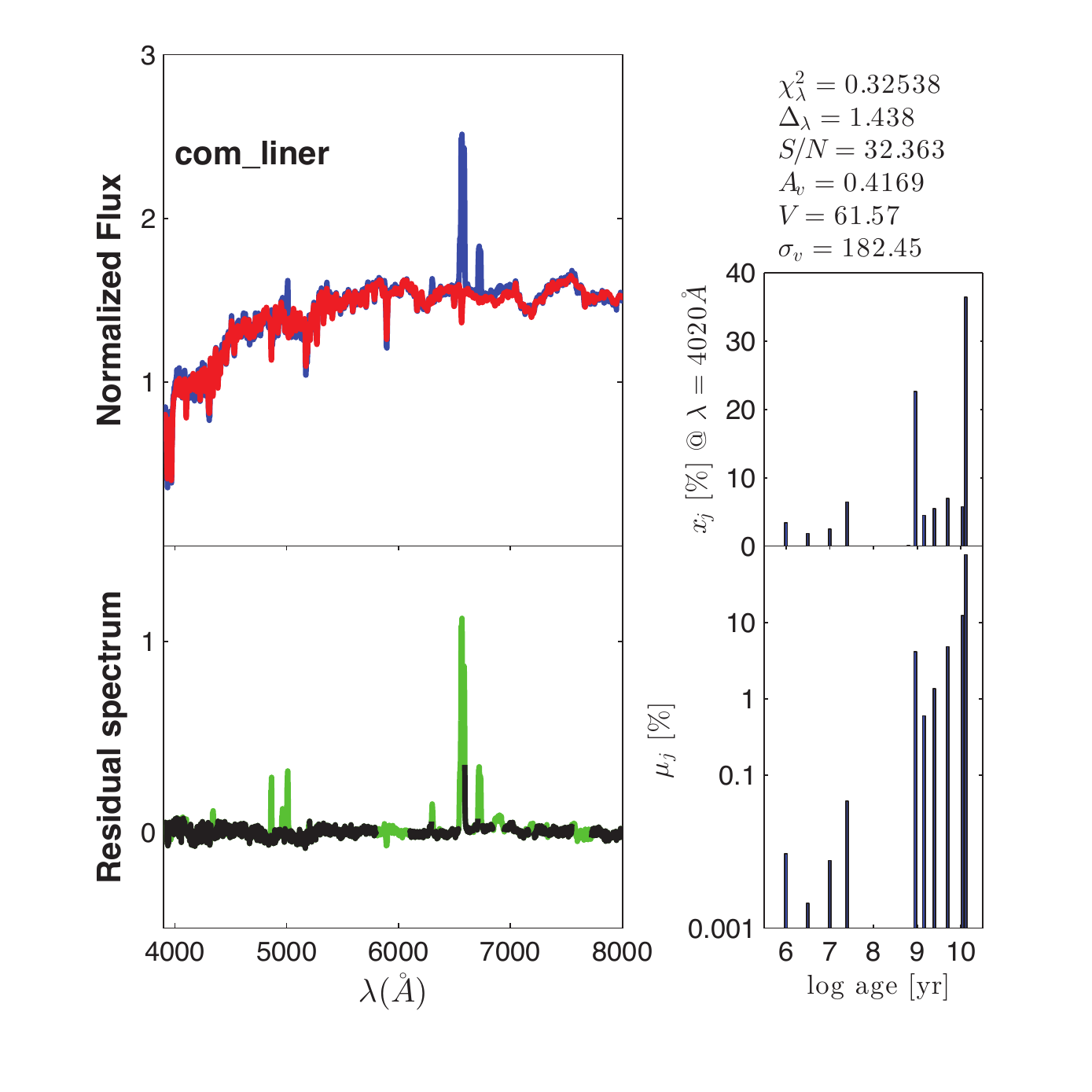}
\includegraphics[width=0.48\textwidth]{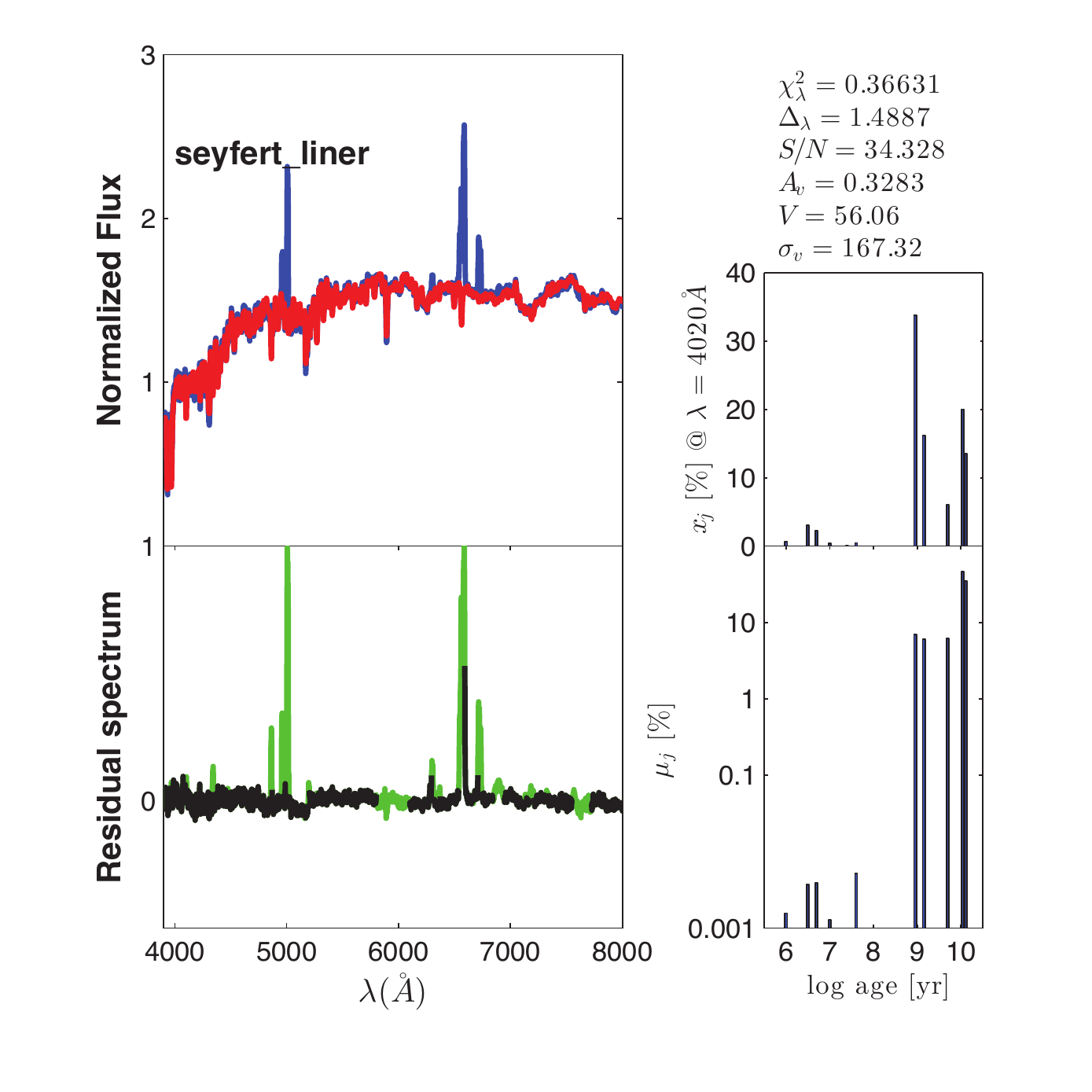}
\caption{Spectral synthesis outputs of the remaining 4 stacked spectra for double-peaked emission-line galaxies from SDSS sample published in \cite{2012ApJS..201...31G}. Each sub-figure shares the same layout as FIGURE \ref{fig7}, and types of the 4 stacked spectra are also tagged on top left of each sub-figure. All symbols are the same as in FIGURE \ref{fig7}.}
\label{fig4}
\end{figure*}

For the double-peaked emission line samples from SDSS DR7 database published in \citet{2012ApJS..201...31G}, and as discussed in SECTION \ref{sect:Sample}, these galaxies are divided into 10 sub-categories according to the locations of their blue-shifted and red-shifted components on BPT diagram. To better illustrate the commonality and property of each sub-sample, we combine all spectra within each subgroup into one stacked spectrum. FIGURES \ref{fig3} and \ref{fig4} display the synthesis outputs for each stacked spectra of 10 subgroups from STARLIGHT, by employing 45 SSPs from BC03, with each sub-figure sharing the same layout and interpretation as FIGURE \ref{fig7}. TABLE \ref{s.ge} also lists the calculated contributions of different types of stellar populations within each subgroup. Since the population synthesis outputs from Dobos' spectra can serve as the regular properties of galaxies from SDSS DR7 database, we analyse their similarities and differences, to learn more about the properties of double-peaked emission-line galaxies. In TABLE \ref{s.ge}, columns 3 to 6 show the synthesis results for double-peaked emission-line objects holding two shifted components with the same BPT diagnosis, we compare the results with the counterpart ones listed in TABLE \ref{s.Dobos}. As for the contributions from stars of different ages, we can see that there still exists a trend on the decrease of young population's significance from 2-SF, 2-COM, 2-Seyfert 2s to 2-LINERs, and 2-LINERs hold the most `old' population contribution, appearing the similar distribution as it shows in Dobos' stacked spectra. Besides these, we also notice some differences, the predominate population of ages in single-peaked SF galaxy spectrum is young population, while in the stacked 2-SF spectrum, it turns out to be the intermediate-age population, with a remarked percentage $\sim$52\%. Although displaying a similar trend, the contribution from young population for each subgroup in Ge's sample is obvious less than its counterpart in Dobos' results. For the stacked spectrum of all 2-Seyfert 2s, which are the most probably dual AGN candidates, the fractional contributions from intermediate-age and old population increase significantly to a total of $\sim$82\%, revealing a more vividly nuclear activities in this kind of sample. Considering the metallicity effects, the dominated contribution remains unchanged within subgroups 2-SF and 2-COM, while in subgroups 2-Seyfert 2s and 2-LINERs, the main contributions come from population with metallicity 2.5 $Z_{\odot}$. Then, we analyse results from spectra hosting components of different BPT types. Although the analysis would be susceptible to the influence of various proportions of different components, we still notice an apparent trend, the sample which integrates one component of SF galaxy type, is obviously rich in young and intermediate-age populations (from $\sim$69.4\% to $\sim$83.8\%), and the sample which holds one component of LINERs type (expect for the SF+LINERs subgroup), would host impressive contribution from old population (from $\sim$32.3\% to $\sim$41.3\%), with an obviously larger value than it in other combined subgroups. The contribution of young populations in each combined subgroup appears to be obvious less that its counterpart in Dobos' stacked spectra, indicating that the double-peaked emitting phenomena is more inclined to happen in an `older' stellar environment. We also find that the subgroups with different BPT types (except for Seyfert 2s+LINERs) would show a significant contribution from populations with metallicities 0.2 and 2.5 $Z_{\odot}$, which suggests that the star formation history of these double-peaked emission line sample is remarkably heterogeneous: young starbursts and old stellar populations all appear in significant and widely varying proportions. For all subgroups that contain one Seyfert II or LINER component, a power-law is considered when carrying out the fittings, and the contributions of this featureless continuum component in different subgroups range from $\sim$4.8\% to $\sim$17.8\%.

\subsection{Results And Comparisons Based On LAMOST Data Set}
\label{sect:LAMOST results}

\begin{figure*}
\centering
\includegraphics[width=0.48\textwidth]{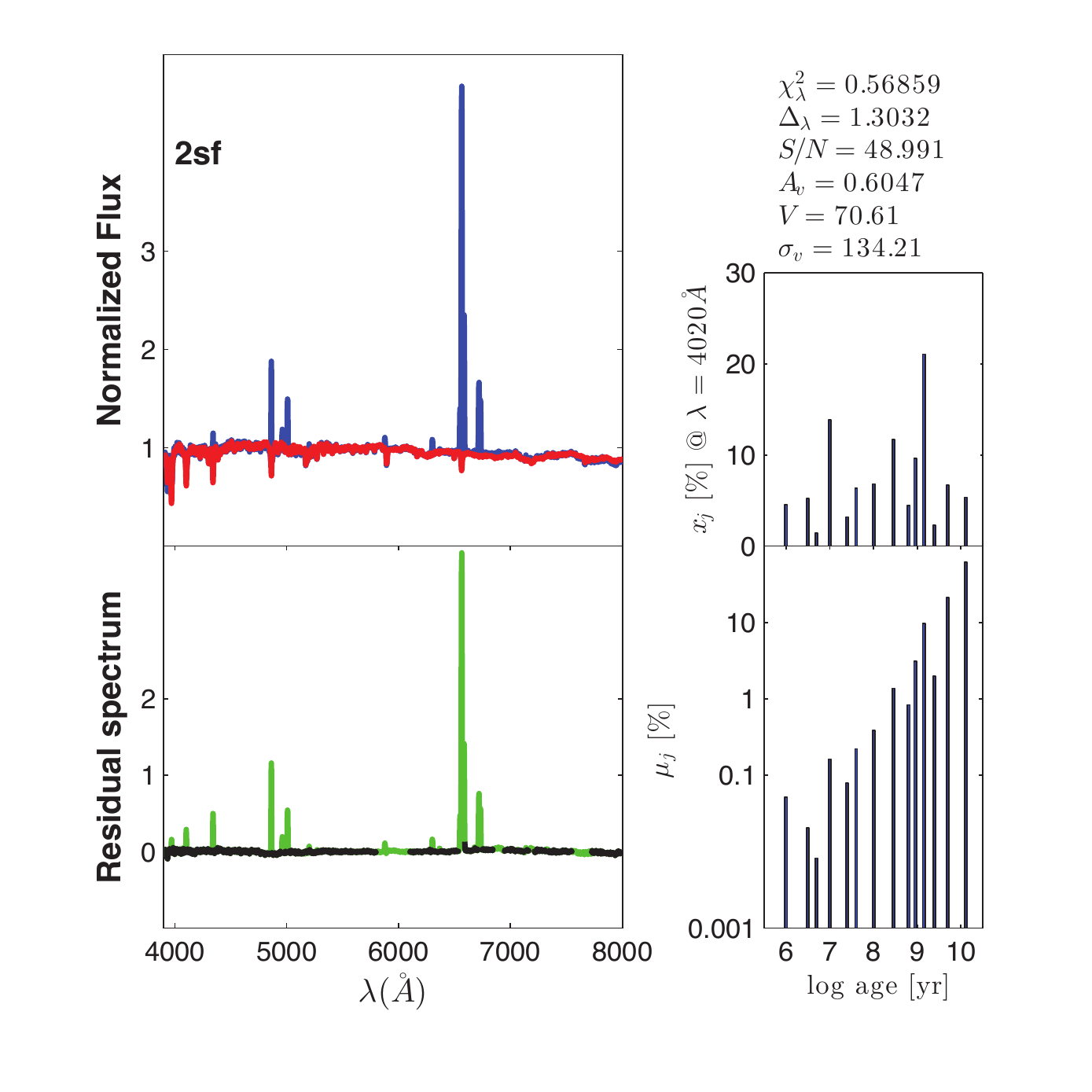}
\includegraphics[width=0.48\textwidth]{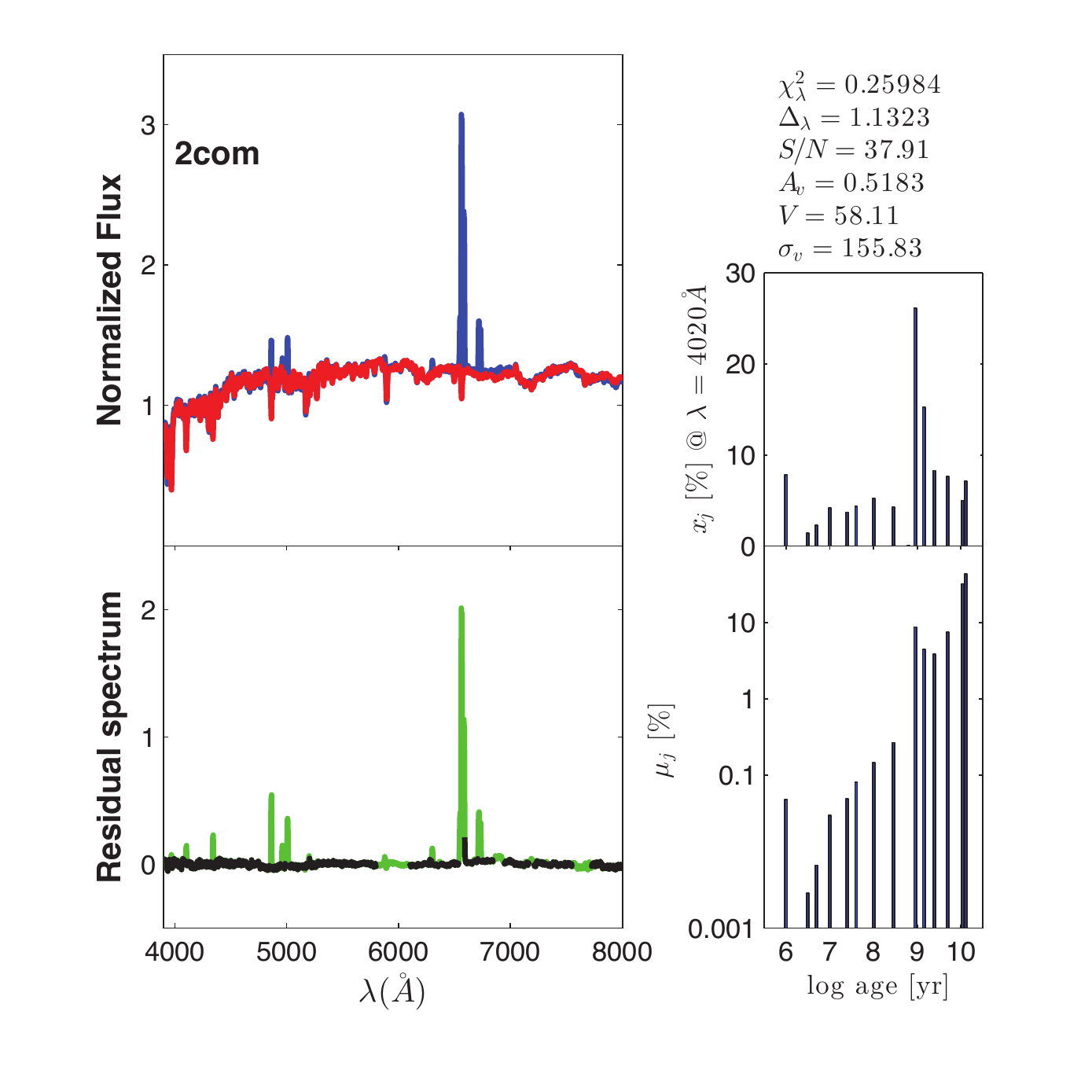}
\includegraphics[width=0.48\textwidth]{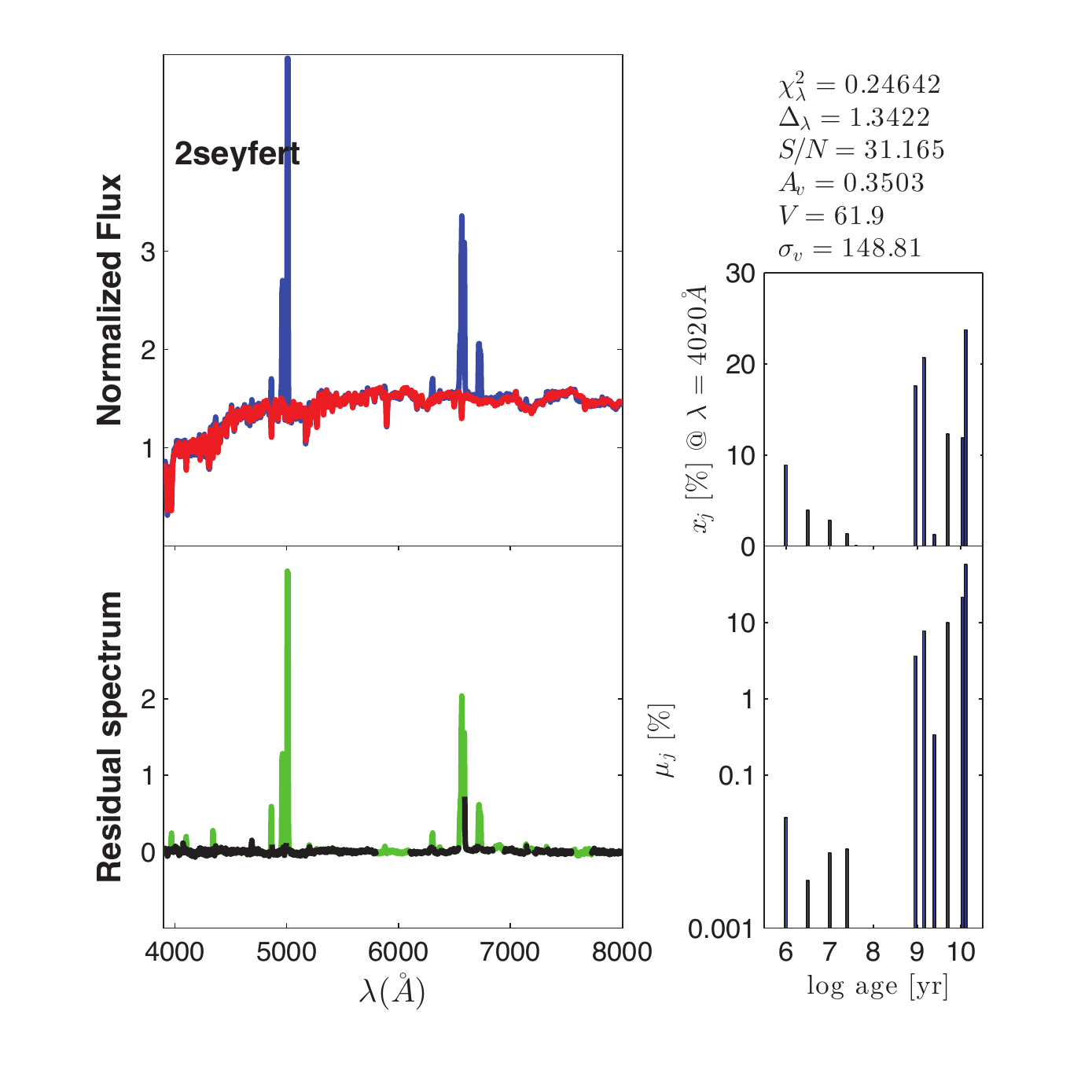}
\includegraphics[width=0.48\textwidth]{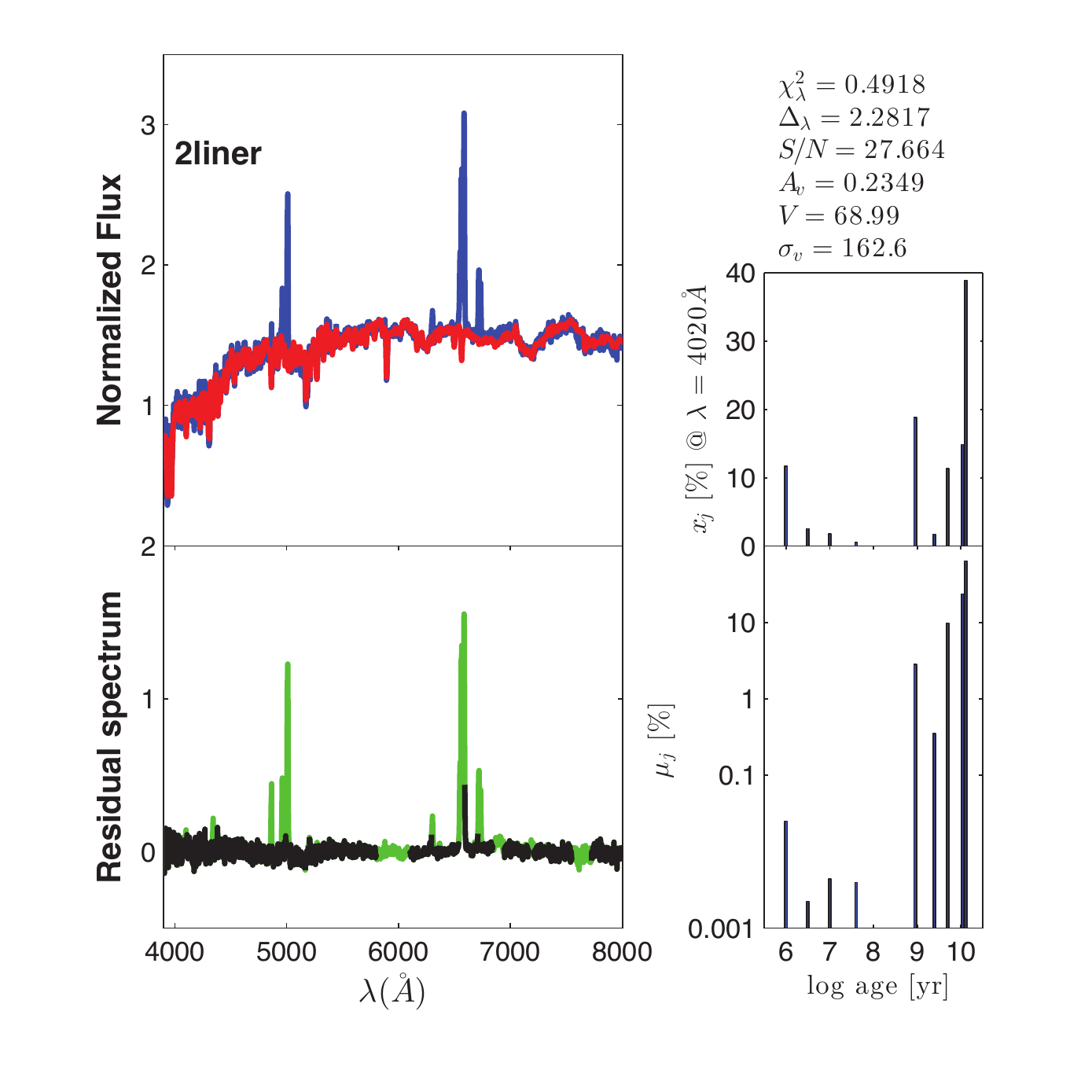}
\includegraphics[width=0.48\textwidth]{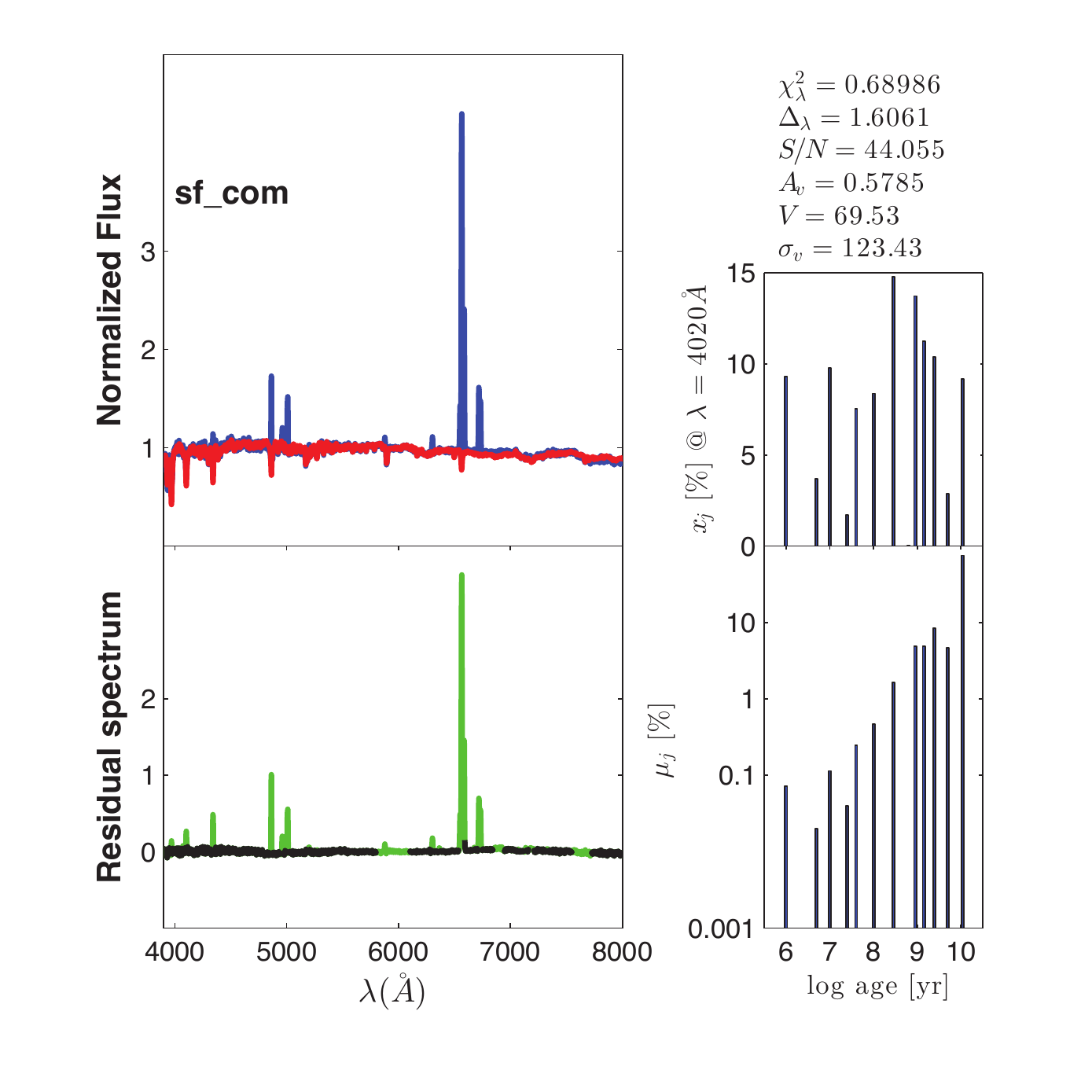}
\includegraphics[width=0.48\textwidth]{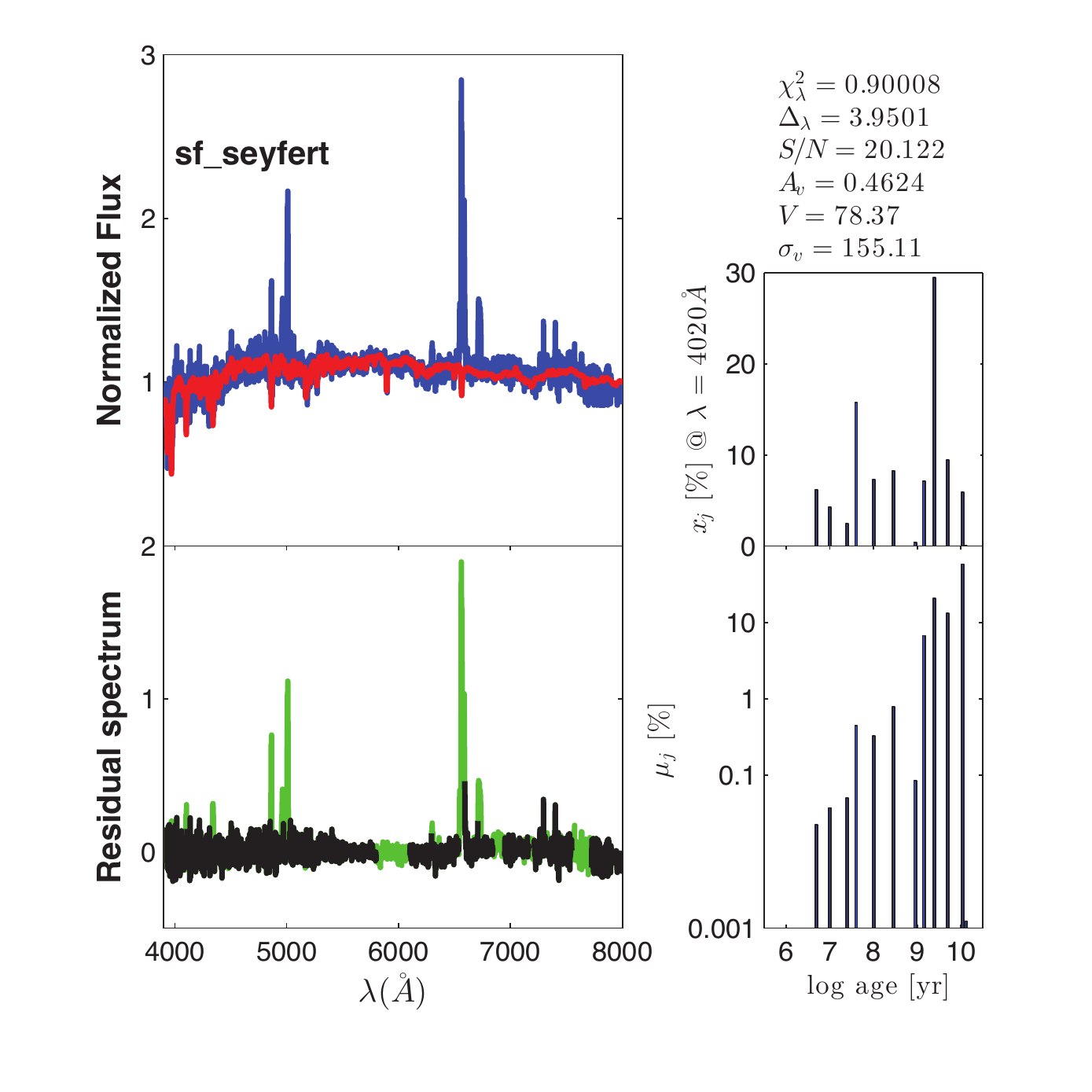}
\caption{Spectral synthesis outputs of 6 stacked spectra for double-peaked emission-line galaxies from the LAMOST sample published in \citet{2019MNRAS.482.1889W}. Each sub-figure shares the same layout as FIGURE \ref{fig7}, and types of the 6 stacked spectra are also tagged on top left of each sub-figure. All symbols are the same as in FIGURE \ref{fig7}.}
\label{fig5}
\end{figure*}

\begin{figure*}
\centering
\includegraphics[width=0.48\textwidth]{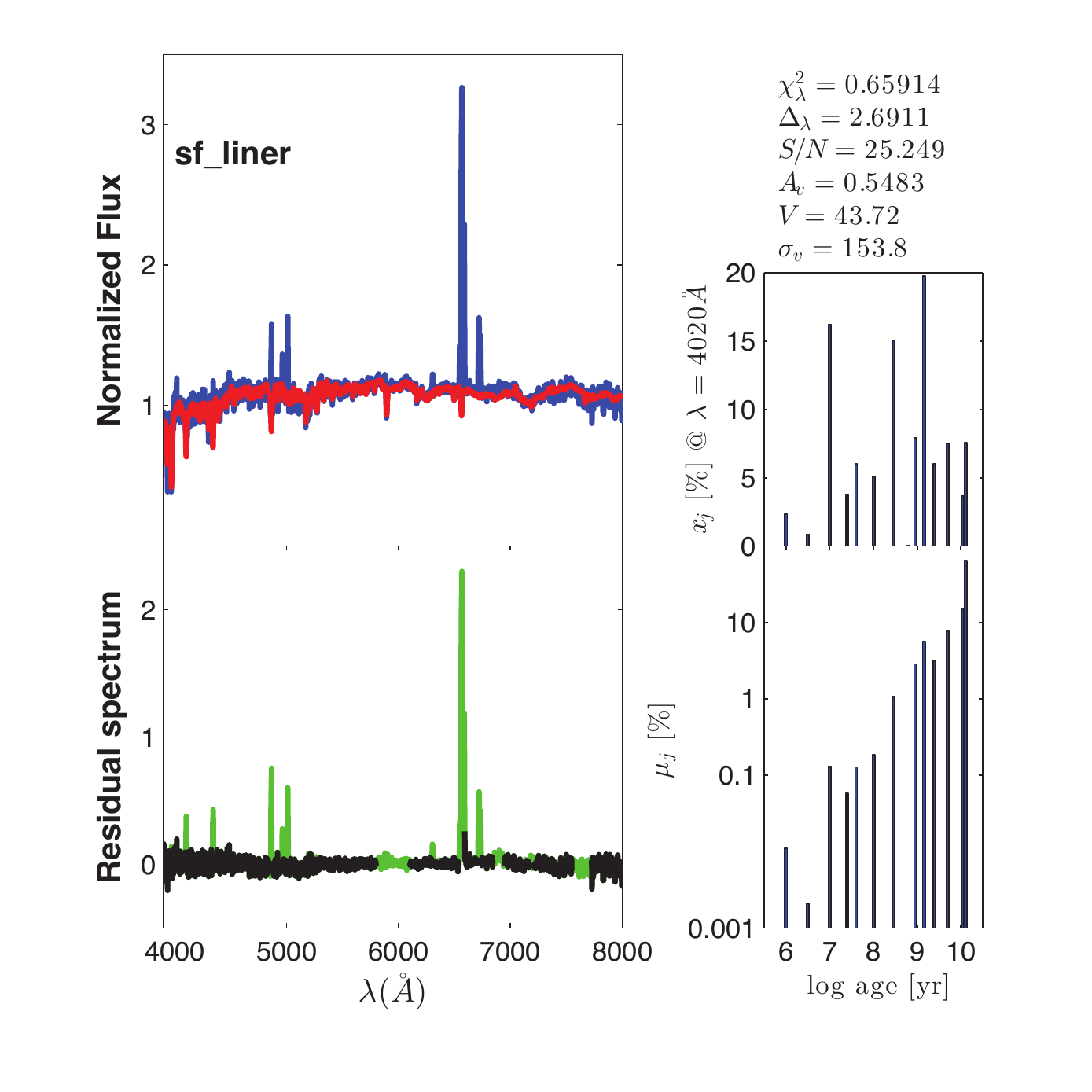}
\includegraphics[width=0.48\textwidth]{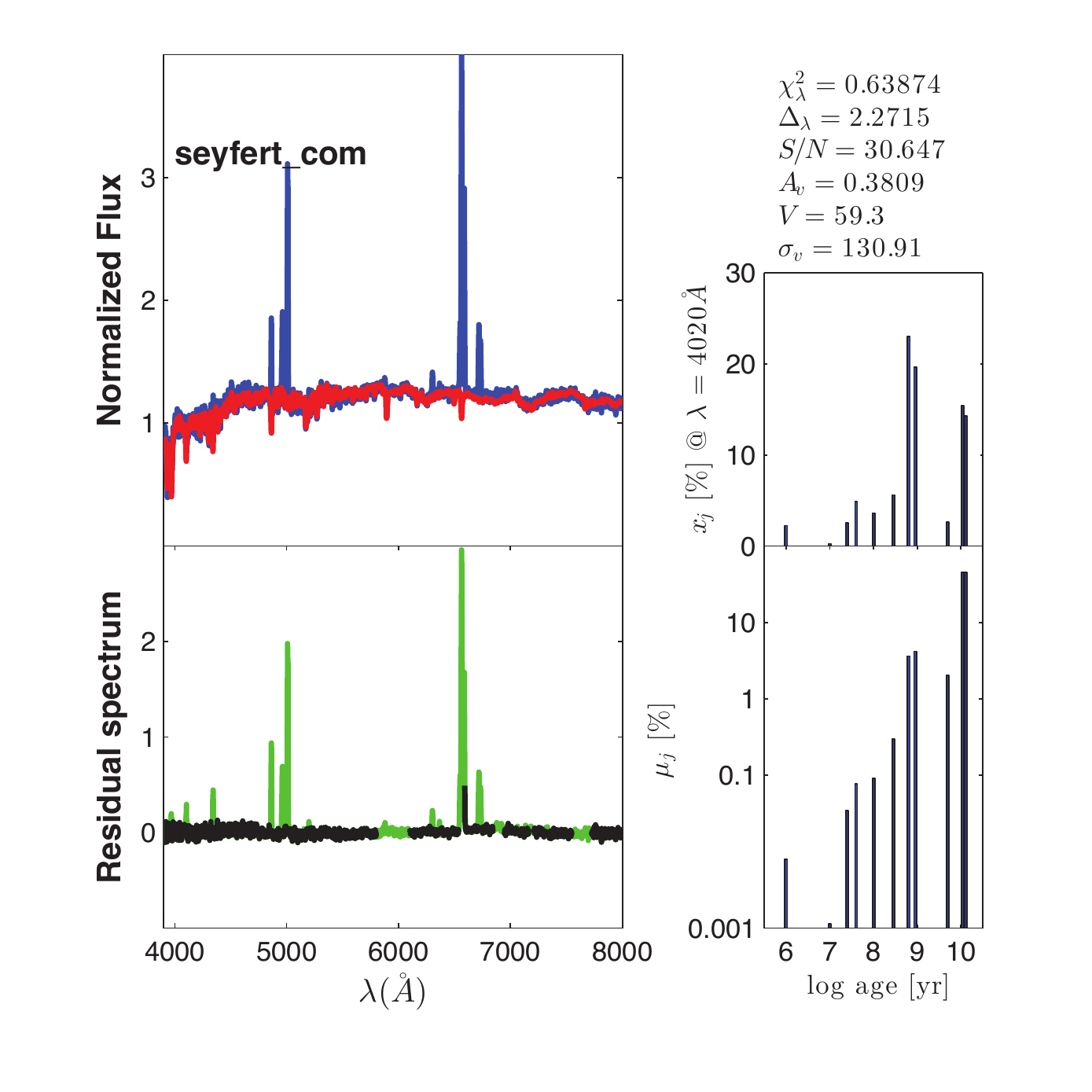}
\includegraphics[width=0.48\textwidth]{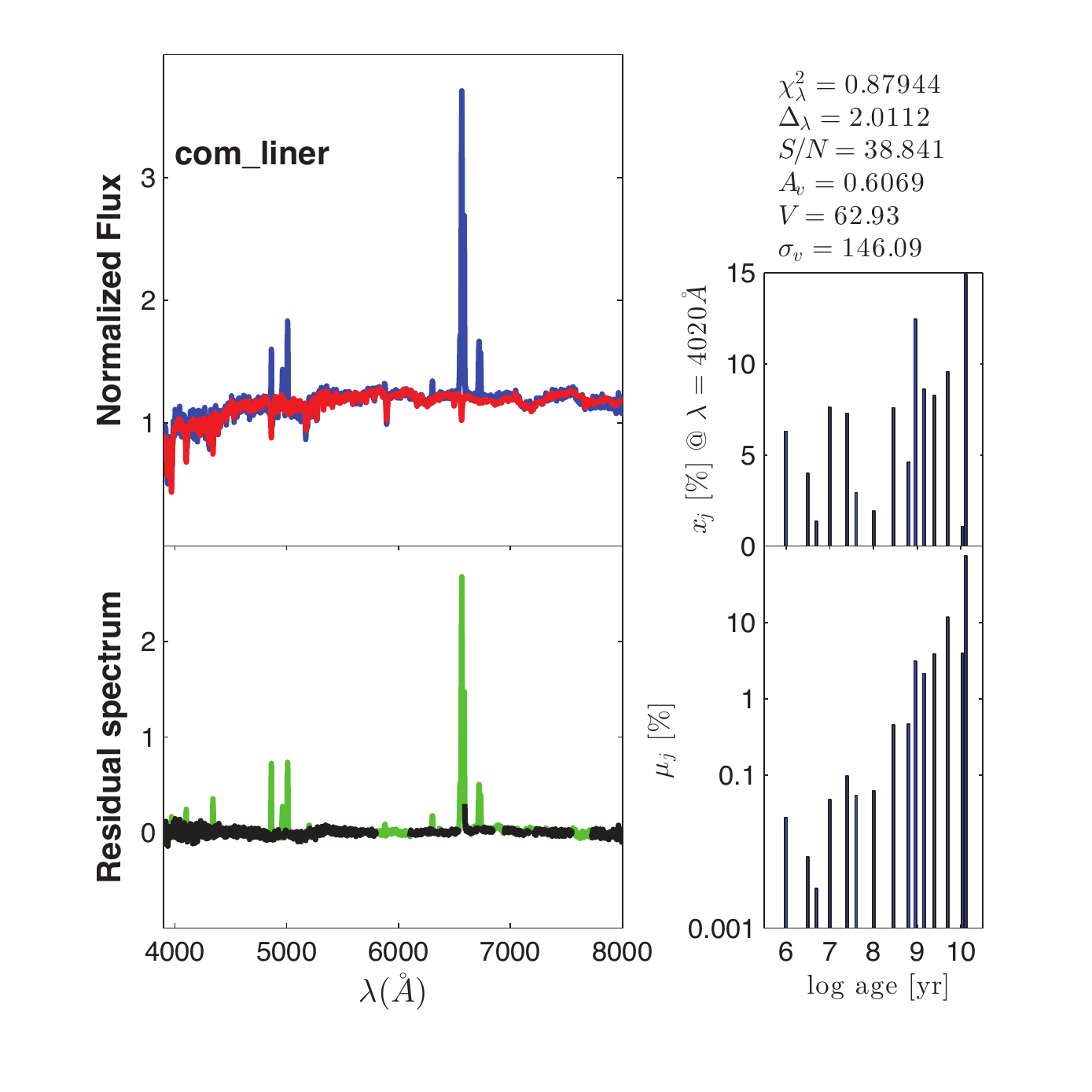}
\includegraphics[width=0.48\textwidth]{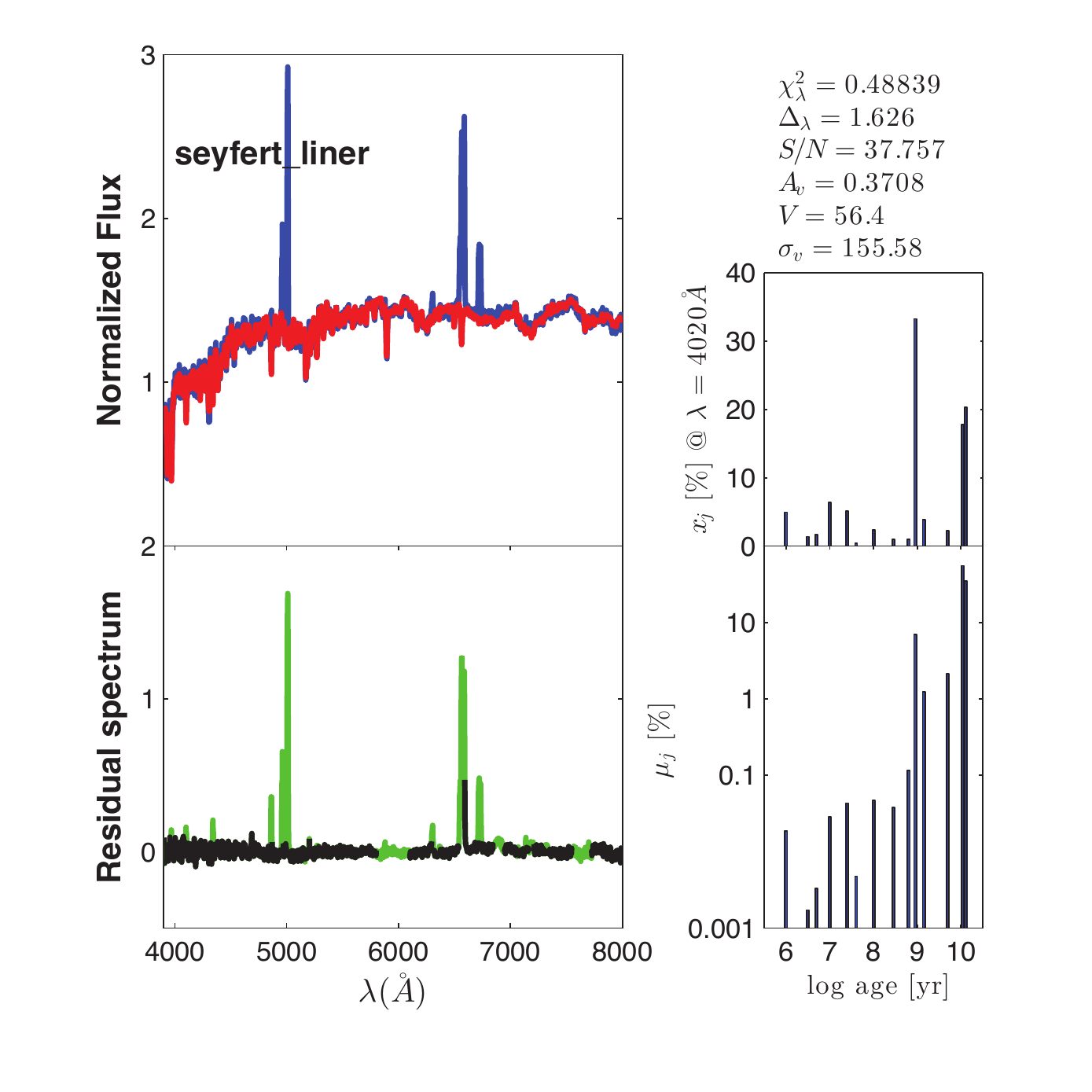}
\caption{Spectral synthesis outputs of the remaining 4 stacked spectra for double-peaked emission-line galaxies from the LAMOST sample published in \citet{2019MNRAS.482.1889W}. Each sub-figure shares the same layout as FIGURE \ref{fig7}, and types of the 4 stacked spectra are also tagged on top left of each sub-figure. All symbols are the same as in FIGURE \ref{fig7}.}
\label{fig6}
\end{figure*}

We also employ STARLIGHT on stacked spectra generated by \cite{2018MNRAS.474.1873W} and \citet{2019MNRAS.482.1889W} from LAMOST database, still using 45 SSPs with the 15 representative ages and 3 metallicities as mentioned in SECTION \ref{sect:SDSS results}. In APPENDIX \ref{sect:appendix}, FIGURES \ref{fig11} to \ref{fig14} illustrate the spectral fits obtained for the stacked spectra provided by \cite{2018MNRAS.474.1873W} drawn from LAMOST DR4 database, and TABLE \ref{s.wll} lists the normalized fractional contributions of SSPs with three bins of ages and three grids of metallicities for 4 stacked spectra from \cite{2018MNRAS.474.1873W}, calculated from STARLIGHT output parameter $x_{j}$. As we can see from this table, the synthesis outputs display the approximately same stellar populations' contribution trend as it is in Dobos' stacked spectra. Except that in view of metallicity effect, the dominated contributions to SF galaxies and AGN + [H~{\sc ii}] come from population with metallicity $Z_{\odot}$, while it appears to be 0.2 $Z_{\odot}$ in Dobos' results. FIGURES \ref{fig5} and \ref{fig6} display the synthesis outputs for 10 stacked spectra constructed from double-peaked emission-line galaxies in LAMOST DR4, and TABLE \ref{s.wmx} lists the percentages for light fractions given by the synthesis outputs. We still notice the decrease of young population's significance from types 2-SF, 2-COM, 2-Seyfert 2s to 2-LINERs, which has been shown in stacked spectra from \citet{2012ApJS..201...31G}. We compare the percentages with the counterpart values listed in TABLE \ref{s.wll}, the significance of young populations in 2-SF, 2-COM and 2-Seyfert 2s are relatively lower, with an average percentage $\sim$7.5\%. We can see that this phenomena is consistent with the conclusion drawn from the similar comparison between Ge and Dobos' synthesis results in SECTION \ref{sect:SDSS results}. In view of stacked spectra holding different BPT components, the subgroups consisting of SF component are obviously rich in young and intermediate-age populations, with the percentages ranging from 88.8\% to 91.1\%, while the ones consisting of Seyfert 2s or LINERs component would be abundant in intermediate-age and old populations, with the percentages ranging from 58.1\% to 71.7\%. Here the contributions of power-law component to different subgroups range from $\sim$0.8\% to $\sim$10.1\%. 

\begin{landscape}

\begin{table*}
\caption{The Quantified Stellar Populations of the Stacked Spectra of Double-peaked Emission-line galaxies from the SDSS Sample Published in \cite{2012ApJS..201...31G}.}
\label{s.ge} \centering
\begin{tabular}{c|c|c|c|c|c|c|c|c|c|c|c}
\hline\hline \multicolumn{2}{c|}{SSP}
  & \multicolumn{10}{c}{Subgroups from Emission-line Diagram}\\
\hline
  &  & 2-SF & 2-COM & 2-Seyfert 2s & 2-LINERs & SF+COM
&SF+Seyfert 2s&SF+LINERs&Seyfert 2s+COM&COM+LINERs&Seyfert 2s+LINERs\\
\hline
age& young &35.0	&	28.4	&	7.5	&	7.4	&	32.5	&	27.3	&	14.8	&	6.5	&	13.7	&	6.5		\\
&intermediate&52.4	&	58.1	&	52.2	&	48.7	&	51.3	&	42.1	&	60.1	&	62.3	&	38.6	&	54.1	\\
&old&12.6	&	13.5	&	29.9	&	39.1	&	16.2	&	12.8	&	16.2	&	13.6	&	41.3	&	32.3	\\
&power law& & & 10.4&4.8&&17.8&8.9&17.6&6.4&7.1\\ 
\hline
$Z/Z_{\odot}$ &0.2 &55.2	&	44.3	&	16.6	&	3.7	&	50.5	&	63.4	&	39.4	&	23.7	&	38.1	&	9.8\\
&1.0 &25.7	&	20.7	&	25.0	&	45.1	&	28.5	&	14.0	&	17.1	&	14.6	&	26.9	&	30.8	\\
& 2.5&19.1	&	35.0	&	48.0	&	46.4	&	21.0	&	4.8	&	34.6	&	44.1	&	28.6	&	52.3\\
&power law & & & 10.4&4.8&&17.8&8.9&17.6&6.4&7.1\\
\hline
\end{tabular}
\end{table*}

\begin{table*}
\caption{The Quantified Stellar Populations of the Stacked Spectra of Double-peaked Emission-line galaxies from the LAMOST Sample Published in \citet{2019MNRAS.482.1889W}.}
\label{s.wmx} \centering
\begin{tabular}{c|c|c|c|c|c|c|c|c|c|c|c}
\hline\hline \multicolumn{2}{c|}{SSP}
  & \multicolumn{10}{c}{Subgroups from Emission-line Diagram}\\
\hline
  &  & 2-SF & 2-COM & 2-Seyfert 2s & 2-LINERs & SF+COM
&SF+Seyfert 2s&SF+LINERs&Seyfert 2s+COM&COM+LINERs&Seyfert 2s+LINERs\\
\hline
age& young &51.7	&	32.4	&	15.4	&	14.5	&	53.8	&	43.3	&	48.5	&	18.2	&	38.1	&	22.1 \\
&intermediate&43.1	&	55.8	&	49.7	&	31.2	&	37.3	&	45.5	&	40.5	&	43.3	&	42.5	&	38.6\\
&old&5.2	&	11.8	&	34.1	&	52.7	&	8.9	&	5.8	&	11.0	&	28.4	&	15.6	&	36.5\\
&power law & & & 0.8& 1.6&&5.4&0.0&10.1&3.8&2.8\\
\hline
$Z/Z_{\odot}$ &0.2 &61.0	&	45.3	&	32.0	&	29.0	&	52.0	&	64.2	&	43.9	&	27.8	&	24.7	&	29.4\\
&1.0&32.5	&	21.7	&	24.1	&	39.0	&	36.9	&	15.4	&	43.3	&	32.1	&	24.1	&	24.7\\
& 2.5&6.5	&	33.0	&	43.1	&	30.4	&	11.1	&	15.0	&	12.8	&	30.1	&	47.4	&	43.1\\
&power law & & & 0.8& 1.6&&5.4&0.0&10.1&3.8&2.8\\
\hline
\end{tabular}
\end{table*}

\end{landscape}

\section{Summary}
\label{sect:summary}

This study focuses on the stellar population physics of double-peaked emission-line galaxies, which has long been perceived as objects related with merging galaxies or other phenomena of disturbed dynamical activities in NLRs.  We group and stack spectra of the double-peaked emitting galaxies published in \cite{2012ApJS..201...31G} and \citet{2019MNRAS.482.1889W}, which are selected from SDSS DR7 and LAMOST DR4 database, respectively. In this work, we fit each double-peaked emission-line sample with mmulti-guassian profiles and classify it according to the BPT types of the blue-shifted and red-shifted components. For each class, the spectra have the same assembled pair of BPT type components. Then, we stack spectra of each class for further study. We also choose the stacked single-peaked emission-line galaxies published in \cite{2012MNRAS.420.1217D} and \cite{2018MNRAS.474.1873W} as the control sample. The single-peaked spectra are also picked out from SDSS DR7 and LAMOST DR4 database and categorized based on BPT types. We fit the continua and spectral absorptions of stacked spectra by using STARLIGHT and 45 SSPs from BC03, and then compare the synthesis results of double-peaked emission-line samples with that of single-picked counterparts, from SDSS and LAMOST databases, independently. From the analysis of the emitted light, we find that the significance of young population shows a downward trend from 2-SF, 2-COM, 2-Seyfert 2s to 2-LINERs, which is similar with the corresponding reference sample. However, the contribution of young population to light of stacked spectra with double-peaks is less than that of single-peaked spectra, which reveals that the double-peaked emitting phenomena is more inclined to happen in an `older' stellar environment. The subgroups consisting of SF component are obviously rich in young and intermediate-age populations in most cases, while the ones consisting of Seyfert 2s or LINERs component tend to be abundant in intermediate-age and old populations, confirming the strong correlations between stellar populations and its spectral classes. Concerning the metallicity effects, subgroups 2-SF and 2-COM are metal-poor, while for subgroups 2-Seyfert 2s and 2-LINERs, the main contributions come from populations with metallicities $Z_{\odot}$ and 2.5 $Z_{\odot}$. The subgroups with different BPT types usually, although not absolutely, show a significant contribution from populations with metallicities 0.2 and 2.5 $Z_{\odot}$, presenting a heterogeneous feature. This feature suggests a more complicated star formation history encoded in these double-peaked emission-line samples than in corresponding single-peaked emission-line samples.

\begin{acknowledgements}
This work was funded by the National Natural Science Foundation of China (NSFC)
under No.11603042 and the China Scholarship Council. The STARLIGHT project is supported by the Brazilian agencies CNPq, CAPES,and FAPESP and by the France-Brazil CAPES/Cofecub program. We thank the wonderful database from LAMOST and SDSS survey. We thank  \cite{2012ApJS..201...31G}, \cite{2012MNRAS.420.1217D} and \cite{2018MNRAS.474.1873W} for their great job.
\end{acknowledgements}

\appendix                  

\section{Spectral Synthesis Fittings of the control samples}
\label{sect:appendix}
\begin{figure*}
\centering
\includegraphics[width=0.8\textwidth]{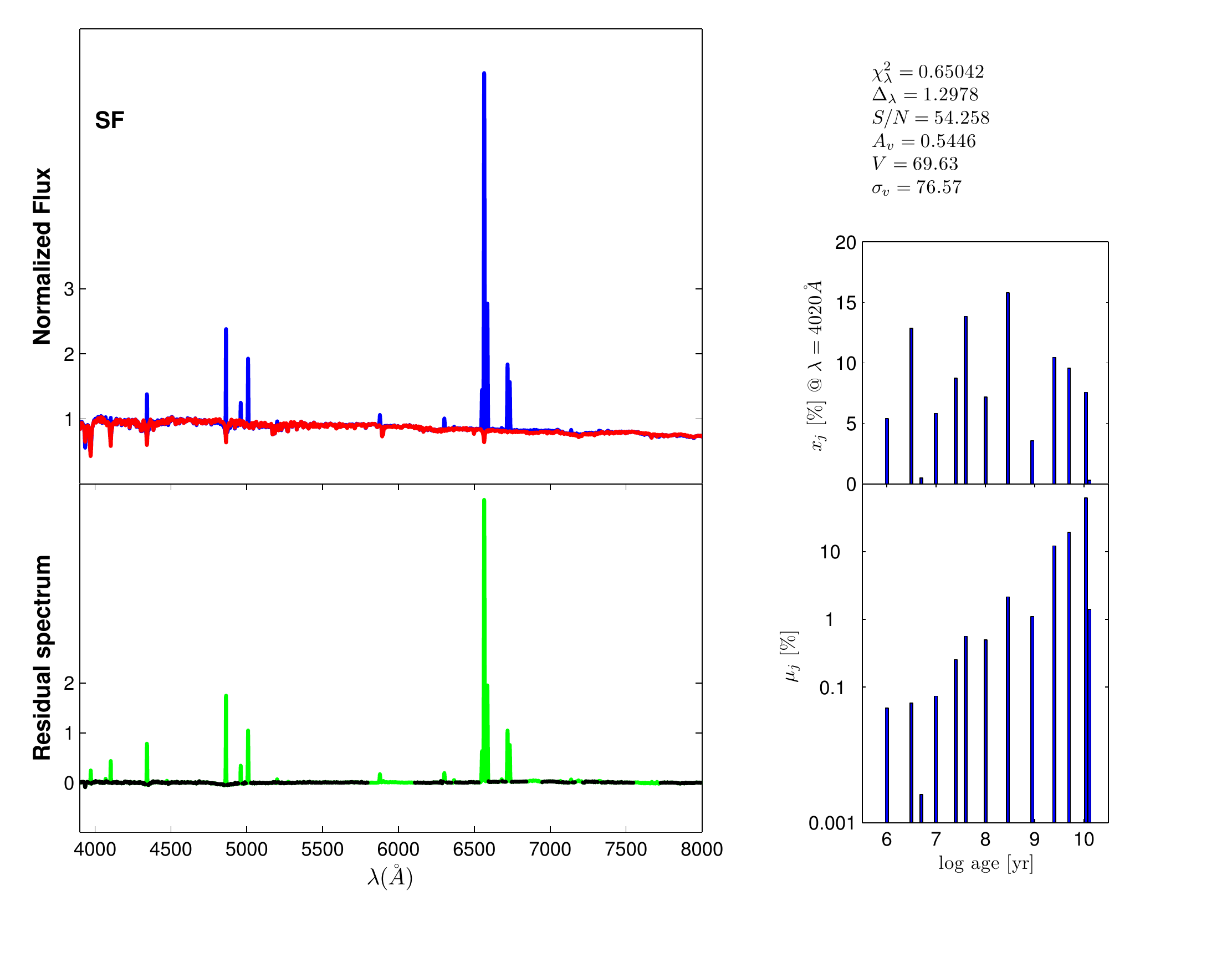}
\caption{Spectral synthesis outputs of the stacked spectrum for SF galaxies within the SDSS sample from \cite{2012MNRAS.420.1217D}. The layout of this figure is as below. Top left: stacked spectrum (blue line),synthesis spectrum (red line). Bottom left: residual spectrum (black line), masked regions (green line). Right: fractions of flux (top) and mass (bottom) as a function of age, in the logarithm coordinate. In top right, some deduced parameters are list, which have been explained in text.}
\label{fig7}
\end{figure*}

\begin{figure*}
\centering
\includegraphics[width=0.8\textwidth]{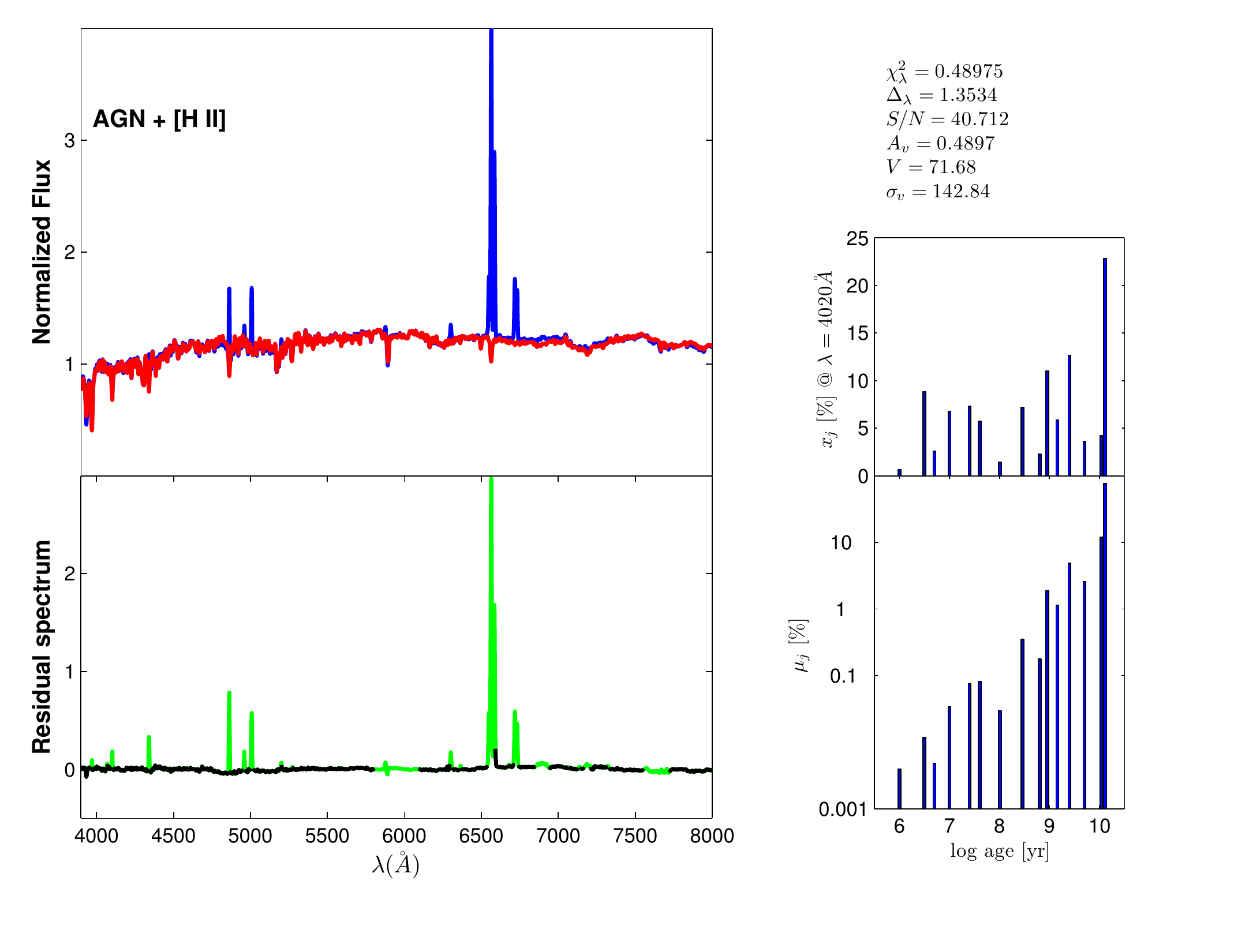}
\caption{Spectral synthesis outputs of the stacked spectrum for AGN + [H~{\sc ii}] galaxies within the SDSS sample from \cite{2012MNRAS.420.1217D}. This figure shares the same layout and symbols as FIGURE \ref{fig7}.}
\label{fig8}
\end{figure*}

\begin{figure*}
\centering
\includegraphics[width=0.8\textwidth]{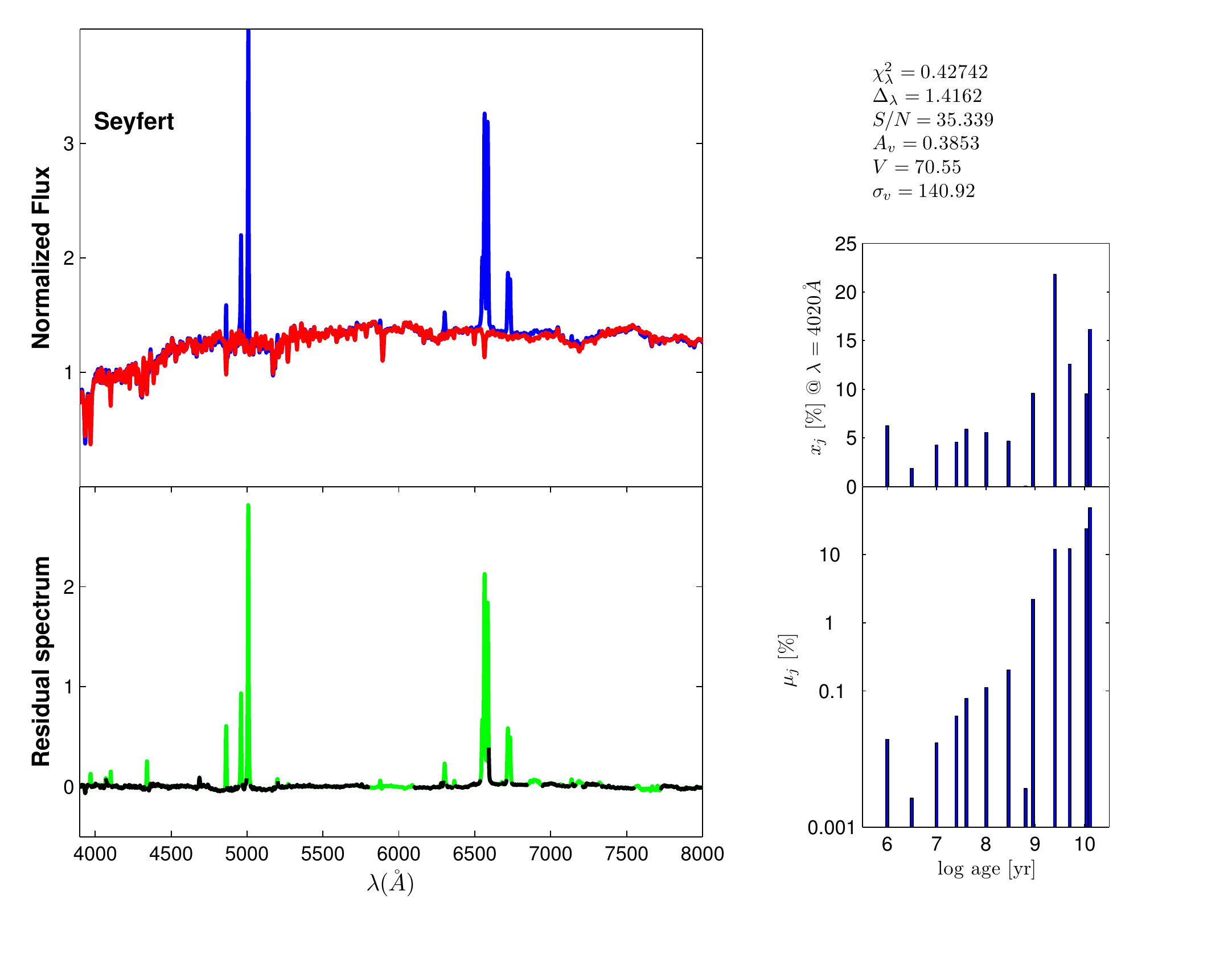}
\caption{Spectral synthesis outputs of the stacked spectrum for Seyfert 2s within the SDSS sample from \cite{2012MNRAS.420.1217D}. This figure shares the same layout and symbols as FIGURE \ref{fig7}.}
\label{fig9}
\end{figure*}

\begin{figure*}
\centering
\includegraphics[width=0.8\textwidth]{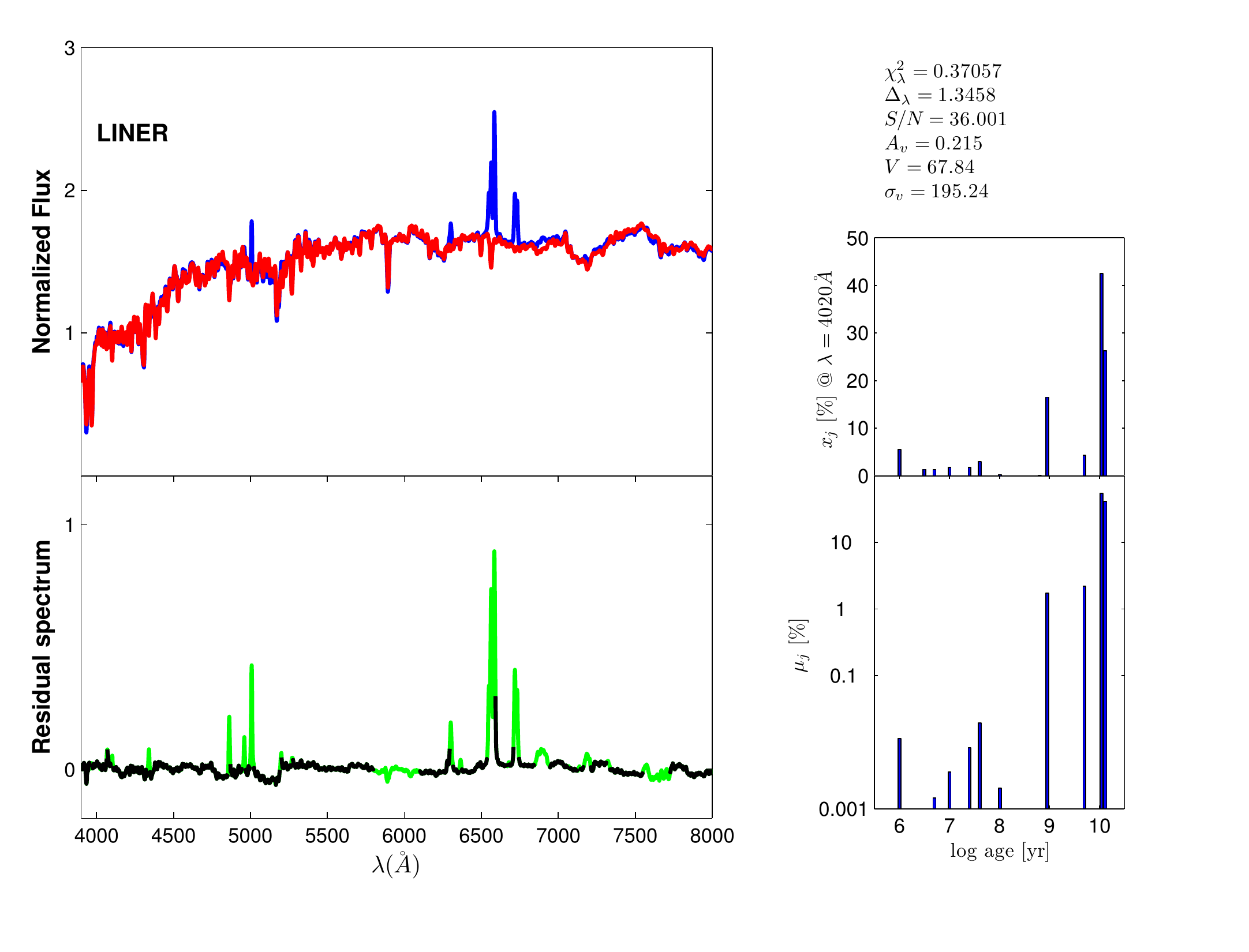}
\caption{Spectral synthesis outputs of the stacked spectrum for LINERs within the SDSS sample from \cite{2012MNRAS.420.1217D}. This figure shares the same layout and symbols as FIGURE \ref{fig7}.}
\label{fig10}
\end{figure*}

FIGURES \ref{fig7} to \ref{fig10} show the synthesis output of the stacked spectra for different types of galaxies within the SDSS sample from \cite{2012MNRAS.420.1217D}, in each figure, the top-left panel displays the stacked spectra (blue line) and the synthesis model (red line), the bottom-left one shows the residual (i.e. the pure emission line spectrum), which is plotted with black line, and the masked regions, which are plotted with a green line, while the two right panels illustrate the deduced star formation history embedded in the population vector $x_{j}$ and $u_{j}$, which are both age-binned in the logarithm coordinate, $x_{j}$ reveals the contribution of each SSP to the model flux at the normalized wavelength $4020\mathrm{\AA}$, while $\mu_{j}$ represents the mass fraction vector. As supplementary information, some derived properties are also listed in the top right. Here are the reduced $\chi^{2}$; $\triangle_{\lambda}$, revealing the difference between observed spectra and synthesis model; $S/N$, which is obtained in a specified region around $4020\mathrm{\AA}$; $A_{v}$, being referred to as the $V$-band extinction; the kinematic parameters, velocity $v$ and the velocity dispersion $\sigma_{v}$. TABLE \ref{s.Dobos} presents the fractional contributions of populations within three quantified bins of ages and three levels of stellar metallicities to the model flux, respectively.

\begin{table*}
\caption{The Quantified Stellar Populations of the Stacked Spectra from the SDSS Sample Published in \citet{2012MNRAS.420.1217D}.}
\label{s.Dobos} \centering
\begin{tabular}{c|c|c|c|c|c}
\hline\hline \multicolumn{2}{c|}{SSP}
  & \multicolumn{4}{c}{Emission-line Diagram}\\
\hline
  &  & SF galaxies& AGN + [H~{\sc ii}]& Seyfert 2s & LINERs\\
\hline
age& young &	69.1	&	39.3	&	31.5	&	13.9	\\
&intermediate&23.2	&	34.4	&	42.8	&	19.9	\\
&old&	7.7	&	26.3	&	25.0	&	65.9	\\	
&power law & & &  0.7& 0.3\\
\hline
$Z/Z_{\odot}$ &0.2 &55.3	&	47.8	&	27.1	&	17.8\\
&1.0 &44.0	&	41.4	&	63.5	&	66.4\\
& 2.5&0.7	&	10.8	&	8.7	&	15.5	\\
&power law & & & 0.7& 0.3\\
\hline
\end{tabular}
\end{table*}

\begin{figure*}
\centering
\includegraphics[width=0.8\textwidth]{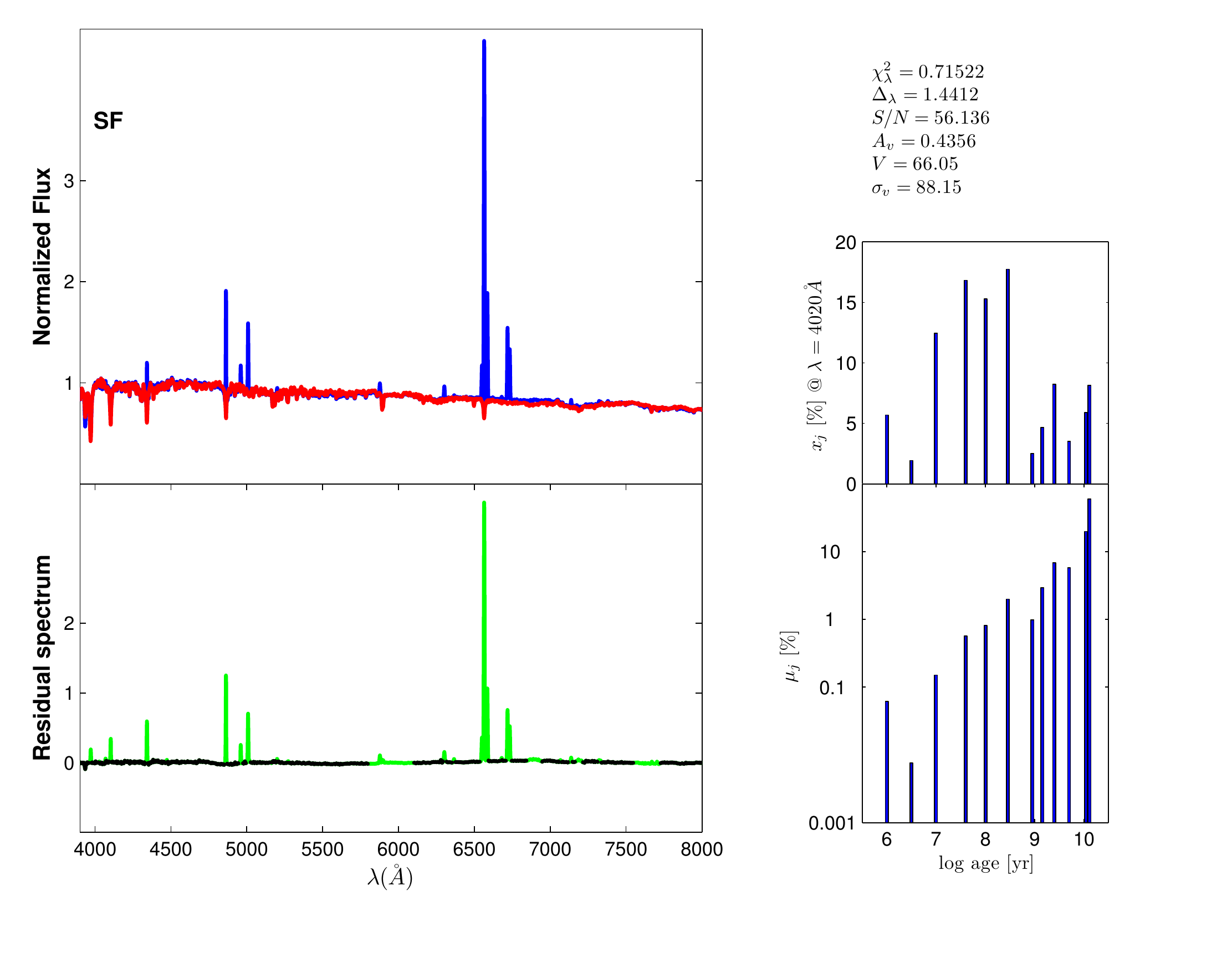}
\caption{Spectral synthesis outputs of the stacked spectrum for SF galaxies within the LAMOST sample from \cite{2018MNRAS.474.1873W}. This figure shares the same layout and symbols as FIGURE \ref{fig7}.}
\label{fig11}
\end{figure*}

\begin{figure*}
\centering
\includegraphics[width=0.8\textwidth]{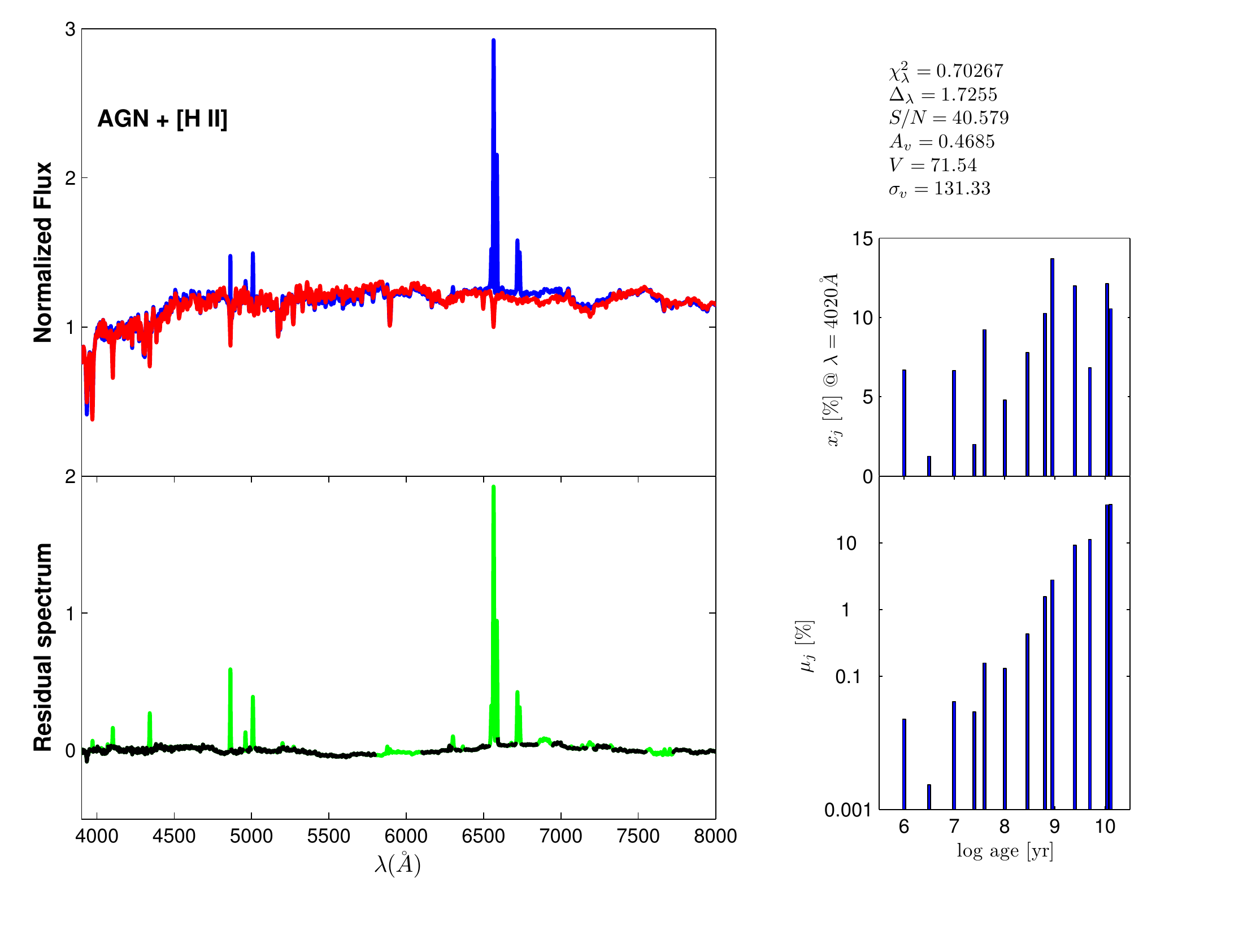}
\caption{Spectral synthesis outputs of the stacked spectrum for AGN + [H~{\sc ii}] galaxies galaxies within the LAMOST sample from \cite{2018MNRAS.474.1873W}. This figure shares the same layout and symbols as FIGURE \ref{fig7}.}
\label{fig12}
\end{figure*}

\begin{figure*}
\centering
\includegraphics[width=0.8\textwidth]{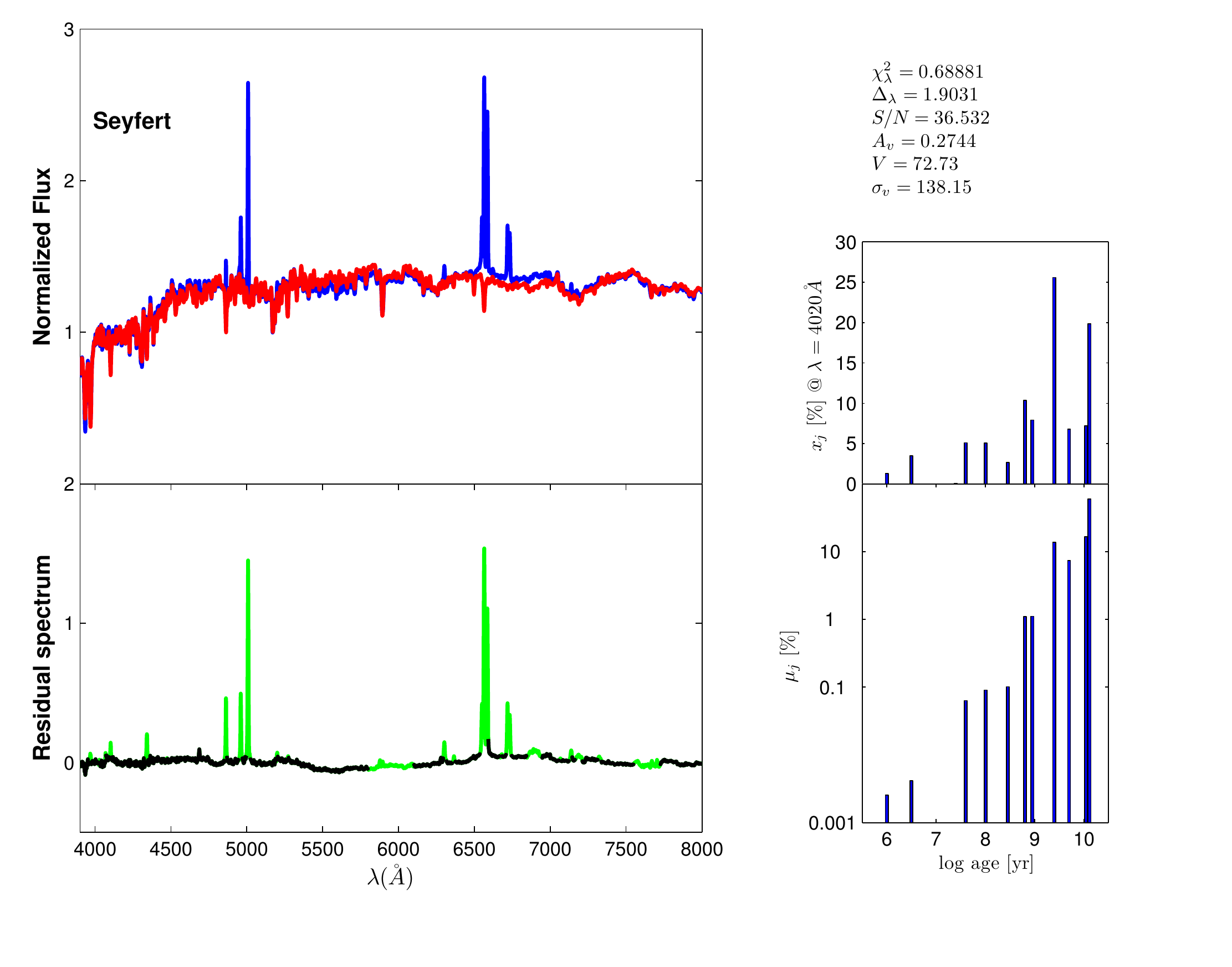}
\caption{Spectral synthesis outputs of the stacked spectrum for Seyfert 2s within the LAMOST sample from \cite{2018MNRAS.474.1873W}. This figure shares the same layout and symbols as FIGURE \ref{fig7}.}
\label{fig13}
\end{figure*}

\begin{figure*}
\centering
\includegraphics[width=0.8\textwidth]{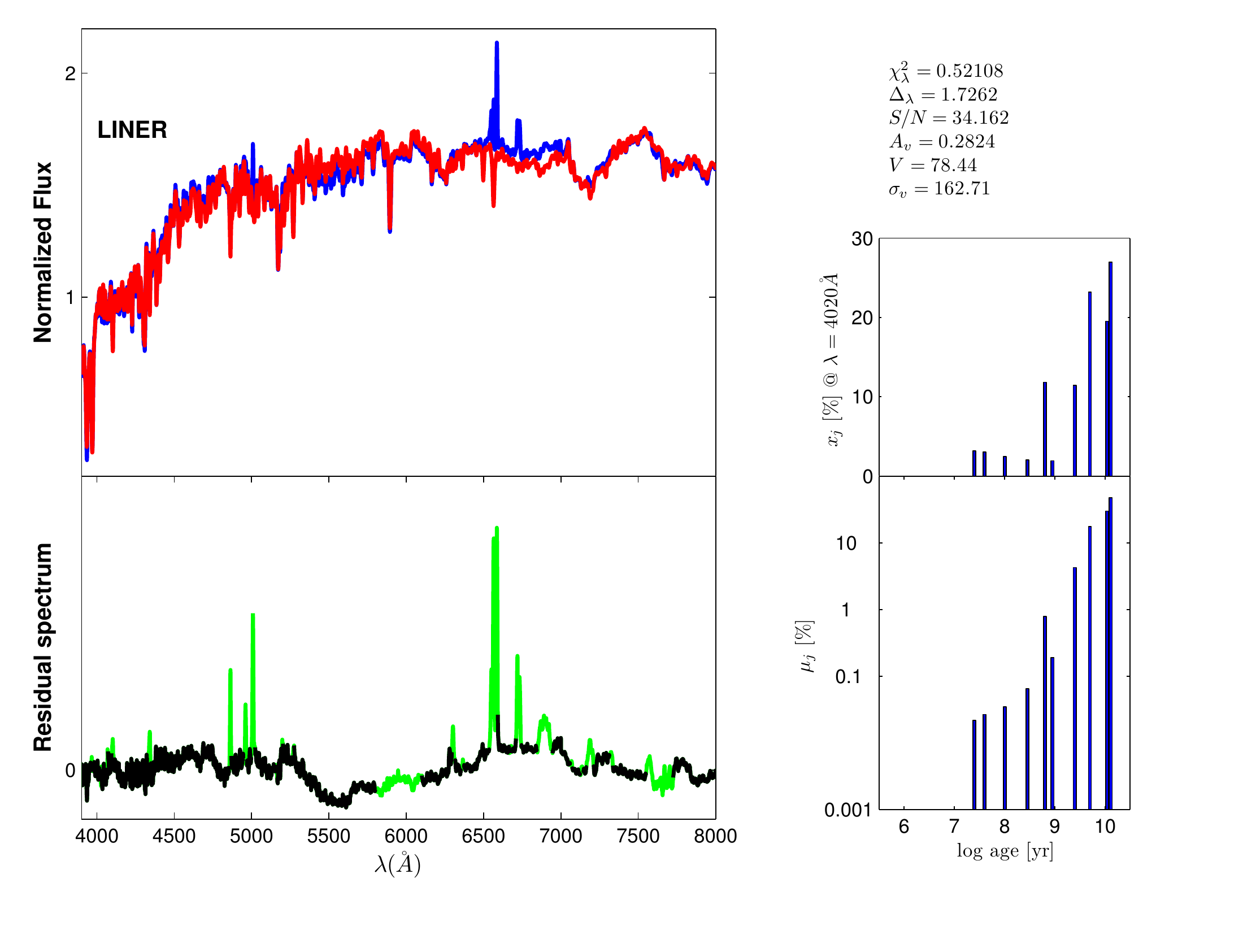}
\caption{Spectral synthesis outputs of the stacked spectrum for LINERs within the LAMOST sample from \cite{2018MNRAS.474.1873W}. This figure shares the same layout and symbols as FIGURE \ref{fig7}.}
\label{fig14}
\end{figure*}

FIGURE \ref{fig11} to \ref{fig14} illustrate the spectral fits obtained for the stacked spectra provided by \cite{2018MNRAS.474.1873W} drawn from LAMOST DR4 database, which can serve as a reference sample to enable us to learn more about the unique characteristics of double-peaked emission-line galaxies. Each figure shares the same layout and symbols as FIGURE \ref{fig7}. TABLE \ref{s.wll} lists the normalized fractional contributions of SSPs with three bins of ages and three grids of metallicities for 4 stacked spectra from \cite{2018MNRAS.474.1873W}, calculated from STARLIGHT output parameter $x_{j}$.

\begin{table*}
\caption{The Quantified Stellar Populations of the Stacked Spectra from the LAMOST sample Published in \citet{2018MNRAS.474.1873W}.}
\label{s.wll} \centering
\begin{tabular}{c|c|c|c|c|c}
\hline\hline \multicolumn{2}{c|}{SSP}
  & \multicolumn{4}{c}{Emission-line Diagram}\\
\hline
  &  & SF galaxies& AGN + [H~{\sc ii}]& Seyfert 2s & LINERs\\
\hline
age& young &68.0	&	37.0	&	17.0	&	10.2\\
&intermediate&18.4	&	41.2	&	48.8	&	43.7\\
&old&13.6	&	21.8	&	26.1	&	44.0\\
&power law & & &  8.2& 2.1\\
\hline
$Z/Z_{\odot}$ &0.2 &39.5	&	19.6	&	8.1	&	5.9\\
&1.0 &60.5	&	71.2	&	70.1	&	83.2\\
& 2.5&0.0	&	9.2	&	13.6	&	8.8\\
&power law & & & 8.2&2.1\\
\hline
\end{tabular}
\end{table*}

\label{lastpage}

\end{document}